\documentclass[11pt, onecolumn, draftcls]{IEEETran}
\usepackage{epsfig, psfrag, amsmath, amssymb, amsfonts, latexsym, eucal}
\newtheorem{lem}{Lemma}
\newtheorem{prop}{Proposition}

\begin{document}

\title{On Outage Behavior of Wideband Slow-Fading Channels}
\author{Wenyi Zhang, {\it Member, IEEE}, and Urbashi Mitra, {\it Fellow, IEEE}
\thanks{The authors are with Communication Sciences Institute, Ming Hsieh Department of Electrical Engineering, University of Southern California, Los Angeles, CA 90089. Email: {\tt \{wenyizha, ubli\}@usc.edu}. This work was supported in part by NSF through grants NRT ANI-0335302, ITR CCF-0313392, and OCE-0520324. The material in this paper was presented in part at the 45th Annual Allerton Conference on Communication, Control, and Computing, Monticello, IL, 2007.}
}

\maketitle

\begin{abstract}
This paper investigates point-to-point information transmission over a wideband slow-fading channel, modeled as an (asymptotically) large number of independent identically distributed parallel channels, with the random channel fading realizations remaining constant over the entire coding block. On the one hand, in the wideband limit the minimum achievable energy per nat required for reliable transmission, as a random variable, converges in probability to certain deterministic quantity. On the other hand, the exponential decay rate of the outage probability, termed as the wideband outage exponent, characterizes how the number of parallel channels, {\it i.e.}, the ``bandwidth'', should asymptotically scale in order to achieve a target outage probability at a target energy per nat. We examine two scenarios: when the transmitter has no channel state information and adopts uniform transmit power allocation among parallel channels; and when the transmitter is endowed with an one-bit channel state feedback for each parallel channel and accordingly allocates its transmit power. For both scenarios, we evaluate the wideband minimum energy per nat and the wideband outage exponent, and discuss their implication for system performance.
\end{abstract}

\newpage

\section{Introduction}
\label{sec:intro}

In wireless communication, it is a well-known fact that spreading the transmit power across a large amount of bandwidths provides an effective means of trading spectral efficiency for power efficiency. A popular trend in next-generation system standards is the proposal of utilizing wideband transmission, in order to boost the system throughput without excess power expenditure in contrast to narrow-band systems. Besides single-carrier or impulse ultra-wideband (UWB)
approaches, a commonly followed design principle is the multicarrier solution like orthogonal frequency division multiplexing (OFDM), which essentially decomposes a frequency-selective wideband channel into a series of frequency-nonselective narrow-band parallel channels, on average each endowed with a small fraction of the total available transmit power. Assuming that the frequency-selective wideband channel models a rich-scattering propagation environment with the number of independent propagation paths growing linearly with the bandwidth, and that the carrier frequencies of parallel channels are sufficiently separated with the effect of the multipath spread essentially eliminated, it is usually a justified approximation that the decomposed narrow-band parallel channels are frequency-nonselective and mutually independent; see, {\it e.g.}, \cite{kennedy69:book}.

Previous theoretical characterizations of wideband channels have primarily focused on the time-varying aspect. The key performance metrics thus are the ergodic achievable information rate and the ergodic capacity, under various channel settings. Furthermore, a highlighted issue there is the lack of precise channel state information (CSI) at the receiver (let alone the transmitter), due to the time-varying nature of the channel model. It has been persistently observed that the lack of precise receive CSI fundamentally affects the efficiency of signaling schemes, in terms of both energy efficiency and spectral efficiency. There has been a vast body of literature in this area, and we refer the reader to \cite{telatar00:it, medard02:it, subramanian02:it, verdu02:it} and references therein and thereafter for major results.

For many practically important scenarios, such as typical indoor channels, however, it has been reported from channel measurement campaigns that the channel response can usually be temporally quasi-static; see, {\it e.g.}, \cite{saleh87:jsac, molisch04:report}. That is, the channel response can be approximated as to remain essentially constant throughout the duration of transmitting an entire coded message. In light of this observation, it is thus not only a theoretically meaningful, but also a practically enlightening, exercise to investigate the wideband transmission problem from a non-ergodic perspective, for which a key performance metric is the outage probability \cite{ozarow94:vt}.

At first glance, the outage probability of wideband systems does not appear to be a well-posed problem. This is because in the wideband limit, the maximum achievable rate, or equivalently the minimum achievable energy per nat, as random variables, generally converge in probability to certain deterministic quantities. More refined characterizations of the outage probability, therefore, need to be studied. To this end, we focus on the exponential decay rate of the outage probability, for a given target energy per nat, as the channel bandwidth measured by the number of parallel channels, increases to infinity. In this paper, we term this exponential decay rate as the wideband outage exponent.

For a fixed operational wideband transmission system, the wideband outage exponent is an incomplete performance descriptor because it does not provide the exact value of the outage probability. However, the wideband outage exponent still provides valuable insights into the system behavior in the regime of (asymptotically) small outage probabilities. Besides its analytical tractability, it serves as a convenient footing for comparing different coding schemes and different channel models. For two wideband transmission systems operating at the same level of energy per nat, if one of them has a wideband outage exponent that is larger than the other's, then we can expect that for sufficiently large bandwidths, the first system achieves smaller outage probabilities.

Our analysis of the wideband outage exponent can be interpreted as to yield two tradeoffs. On the one hand, it reveals a tradeoff between rate and reliability. This type of tradeoff is not entirely a new topic for fading channels.\footnote{Such a tradeoff, in fact, has long been one of the central themes in information theory, as captured by the channel reliability function \cite{gallager68:book}.} Specifically, it has been extensively studied in the context of narrow-band high-power systems with multiple antennas, since the seminal work of \cite{zheng03:it}. There, the rate is quantified by the multiplexing gain and the reliability by the diversity gain, and thus the so-called multiplexing-diversity tradeoff indicates how the reliability is reduced as the rate increases. We note that, the diveristy order is usually another ``outage exponent'', because for typical slow-fading channels it essentially corresponds to the exponential decay rate of the outage probability as the signal-to-noise ratio (SNR) grows large. In contrast, the wideband outage exponent studied in this paper is the exponential decay rate of the outage probability as the number of parallel channels, {\it i.e.}, bandwidth, grows large.

On the other hand, our analysis reveals a tradeoff between energy efficiency and spectral efficiency. From a high-level perspective, the wideband outage exponent for wideband quasi-static fading channels plays a similar role as the wideband slope \cite{verdu02:it} for wideband ergodic fading channels, in that both asymptotically indicate the amount of bandwidths required to support a target energy per nat. The wideband slope essentially corresponds to the relative (compared to the first-order derivative) significance of the second-order derivative of the fading-averaged achievable rate per narrow-band parallel channel, at vanishing power. Thus it captures the loss in energy efficiency due to large, yet still finite, operating bandwidth. The wideband outage exponent essentially corresponds to the outage probability as an exponentially decreasing function of bandwidths, at a target energy efficiency. Thus it captures the required amount of bandwidth for attaining the target energy efficiency, in order to satisfy a given reliability constraint as specified by the outage probability.

In the ergodic fading case, it suffices to separately code for each single parallel channel and thus to convert the wideband problem into a low-power problem; see, {\it e.g.}, \cite{verdu02:it}. This approach is lossless in terms of achievable rate there because it exploits the temporal ergodicity of each parallel channel. In the quasi-static fading case under an outage probability criterion, however, it is necessary to adopt joint coding across all the parallel channels. In contrast, the alternative approach, separate coding for each single parallel channel concatenated with a cross-channel erasure coding stage, can be substantially sub-optimal due to the ``hard thresholding'' effect by outage events for individual parallel channels. In the analysis of the joint coding approach, we utilize some basic results in the theory of large deviations.

In the following, we briefly summarize the main results and observations in this paper.

\begin{enumerate}
\item For the case in which the transmitter has no CSI and adopts uniform transmit power allocation among parallel channels, we derive general formulas for the wideband minimum energy per nat and the wideband outage exponent. It is shown that these quantities depend upon statistics, specifically, the expectation and the logarithmic moment generating function, of only the first-order derivative of the per-channel achievable rate random variable at vanishing SNR.

\item Applying the general formulas, we evaluate the wideband outage exponent for scalar Rayleigh, Rician, and Nakagami-$m$ fading channels, as a function of the target energy per nat. Rician fading channels achieve larger wideband outage exponents than the Rayleigh case, but the gap becomes evident only when the line-of-sight component is dominant. The wideband outage exponent of Nakagami-$m$ fading channels is precisely $m$ times that of the Rayleigh case, implying that in the wideband regime an $m$-fold bandwidth savings can be expected for Nakagami-$m$ fading channels in contrast to Rayleigh fading channels.

\item We then evaluate the wideband outage exponent for multiple-input-multiple-output (MIMO) fading channels, with and without spatial antenna correlation, for spatially white and covariance-shaped circular-symmetric complex Gaussian inputs. For channels with spatially uncorrelated Rayleigh fading and spatially white inputs, the wideband outage exponent is essentially identical to that of the scalar Nakagami-$m$ fading case, with the parameter $m$ equal to the product of the numbers of transmit and receive antennas. For channels with spatial antenna correlation, a covariance-shaping problem needs to be solved for the input covariance matrix to maximize the wideband outage exponent, for the given target energy per nat. In general, the optimal input covariance matrix for minimizing the wideband minimum energy per nat may not be the optimal choice for maximizing the wideband outage exponent, especially as the target energy per nat gets large.

\item We finally examine the case in which the transmitter has partial information of the channel state. We analyze a specific transmission protocol in which the transmitter has an one-bit information for each parallel channel indicating whether the squared fading coefficient (assuming a scalar Rayleigh fading model) is beyond a certain threshold. Based upon such an one-bit feedback, the transmitter adaptively allocates its power among the parallel channels. We obtain the wideband minimum energy per nat and the wideband outage exponent for the transmission protocol. The wideband minimum energy per nat can be made arbitrarily
small, by simply adjusting the protocol parameters such that most (or even all) transmit power is allocated for sufficiently strong parallel channels. The corresponding wideband outage exponent, however, quantifies the penalty in terms of outage probability, as the energy efficiency improves. This tradeoff is inevitable, because in order to achieve a smaller value of the energy per nat reliably, more parallel channels are needed, to statistically ensure that there exist sufficiently many strong parallel channels. When the transmission protocol employs on-off power allocation, allocating transmit power only for sufficiently strong parallel channels, the wideband outage exponent is reduced at an exponential speed as the wideband minimum energy per nat decreases at an inversely-linear speed, implying that the bandwidth needs to scale exponentially, in order to maintain a fixed level of outage probability asymptotically.
\end{enumerate}

In \cite{ray05:ciss}, the authors investigate the scaling of codeword error probabilities in low-power block Rayleigh fading channels with uncorrelated multiple antennas. Their approach examines the codeword error probabilities for one parallel channel across several fading blocks, and simultaneously evaluates the fading-induced outage part and the noise-induced part of decoding errors, finding that the former is usually dominant for typical slow-fading channels. In this paper, our approach explicitly connects the parallel channel power to the bandwidth scaling, and exclusively examines the outage probability, assuming that the noise-induced error probability has been made arbitrarily small by coding over sufficiently long codewords, in light of the quasi-static condition of the fading process. Such a simplification still captures the essence of quasi-static fading channels, and permits the analysis to yield additional insights into the system behavior, under more general settings as summarized above. It is worth noting that in \cite{ray05:ciss}, its {\it ad hoc} analysis for uncorrelated Rayleigh MIMO channels has also identified the gain factor of the wideband outage exponent, as the product of the numbers of transmit and receive antennas.

Throughout this paper, we assume that the receiver possesses perfect CSI and thus can adopt coherent reception. Such an assumption does not appear as a fundamental issue in the context of quasi-static fading channel model, and furthermore, in the low-power (per parallel channel) regime non-coherent reception usually attains first-order equivalence to coherent reception \cite{verdu02:it} thus would lead to the same wideband outage exponent. Regarding the transmit knowledge of CSI, however, we only consider the no-CSI or partial-CSI (specifically, one-bit CSI per parallel channel) case in this paper, and leave the case of transmit full-CSI for future exploration. On one hand, the transmitter of a wideband system is unlikely to be endowed with precise channel state information due to its impractical demand on a high-capacity feedback link, so the case of transmit full-CSI is a less imperative topic for practical considerations. On the other hand, when the transmitter has precise knowledge of CSI, it can thus fully adapt its power as well as rate allocation, under short/long-term power and fixed/variable rate constraints \cite{caire99:it, caire04:it}. A full account of these topics is beyond the scope of this paper and deserves a separate treatment.

The remainder of this paper is organized as follows. Section \ref{sec:general} formulates the general problem and establishes the general formulas of the wideband minimum energy per nat and the wideband outage exponent. Section \ref{sec:scalar} specializes the general results to scalar channels, and Section \ref{sec:mimo} addresses the effects of multiple antennas and antenna correlation. Section \ref{sec:feedback} turns to the partial transmit CSI case, solving the problem for a specific transmission protocol with one-bit channel state feedback per parallel channel. Finally Section \ref{sec:conclusion} concludes this paper. All the logarithms are to base $e$, and information is measured in units of nats.

\section{General Result}
\label{sec:general}

Consider $K$ parallel channels $\mathcal{C}_k$, $k = 1, \ldots, K$. We assume that all the channels have identical statistics, characterized by the transition probability distribution $p(y|x, s)$, where $x \in \mathcal{X}$ is the input, $y \in \mathcal{Y}$ the output, and $s \in \mathcal{S}$ the channel state. $\mathcal{X}$, $\mathcal{Y}$, and $\mathcal{S}$ are appropriately defined measurable spaces. For simplicity, in this paper we further assume that the $K$ parallel channels are mutually independent, and thus the channel states $S_k$, $k = 1, \ldots, K$, are also mutually independent. The quasi-static condition indicates that each $S_k$ remains constant over the entire coding block. So for each coding block, we can interpret $\{S_k\}_{k = 1}^K$ as $K$ independent and identically distributed (i.i.d.) realizations of some channel state distribution $p_S(s)$.

Regarding the channel state information, we assume that the receiver has perfect knowledge of $S_k$, $k = 1, \ldots, K$, but the transmitter has no knowledge of them for lack of an adequate feedback link. In Section \ref{sec:feedback}, we will endow the transmitter with an one-bit channel state feedback for each parallel channel.

By defining a cost function for each channel input $x \in \mathcal{X}$, $c(x) \geq 0$, we can associate with an input distribution $p_X(x)$ an average cost $c_X = \mathbf{E}[c(X)]$. If we impose a total average cost constraint $\rho$, and choose an input distribution satisfying $c_X = \rho/K$ for each parallel channel $\mathcal{C}_k$, $k = 1, \ldots, K$, then from the coding theorem for parallel channels (see, {\it e.g.}, \cite{gallager68:book}), the total achievable rate conditioned upon the channel state realizations as we code jointly over the $K$ parallel channels is
\begin{eqnarray}
R(K, \rho) &:=& \sum_{k = 1}^K I(X_k; Y_k, S_k)\nonumber\\
&=& \sum_{k = 1}^K I(X_k; Y_k|S_k),
\end{eqnarray}
where $X_k$ and $Y_k$ denote the input and output of channel $\mathcal{C}_k$. Here we utilize the fact that the channel input is independent of the channel state, due to the lack of transmit CSI. Note that $X_k$ incurs the average cost $\rho/K$, for $k = 1, \ldots, K$. The per-channel mutual information $I(X_k; Y_k|S_k)$ is a random variable induced by $S_k$, therefore we denote it by $J(\rho/K, S_k)$ in the sequel. The total achievable rate then can be rewritten as
\begin{eqnarray}
R(K, \rho) = \sum_{k = 1}^K J(\rho/K, S_k),
\end{eqnarray}
which is the sum of $K$ i.i.d. random variables. We note that it is definitely legitimate to let parallel channels be allocated different average costs, under the total average cost constraint $\rho$. In this paper, for simplicity, we only consider the uniform cost allocation, except in Section \ref{sec:feedback}, where the one-bit channel state feedback leads to a certain capability of adaptive power allocation. Such a uniform cost allocation is also in spirit with the basic philosophy of wideband systems, which advocate spreading the energy across large bandwidths \cite{simon94:book}.

Letting $K$ grow from one to infinity, and specifying accordingly the input distributions indexed by $K$, we obtain a sequence of achievable rate random variables $\left\{R(K, \rho)\right\}_{K = 1}^\infty$. We then use outage probability to characterize the system performance. We specify a target rate $r \geq 0$ in nats, then the outage probability $\mathcal{O}(K, \rho, r)$ is
\begin{eqnarray}
\mathcal{O}(K, \rho, r) := \mathrm{Pr}\left[R(K, \rho) \leq r\right].
\end{eqnarray}
In this paper, we are primarily interested in the wideband asymptotic behavior of $\mathcal{O}(K, \rho, r)$. By introducing the average cost per nat,\footnote{The corresponding cost per bit can be easily obtained by shifting down the cost per nat by $\ln 2 = -1.59 \mbox{dB}$.}
\begin{eqnarray}
\eta := \frac{\rho}{r},
\end{eqnarray}
we can characterize the wideband asymptotic decay rate of $\mathcal{O}(K, \rho, r)$ by the wideband outage exponent
\begin{eqnarray}
\mathcal{E}(\eta) := \lim_{K \rightarrow \infty} \frac{-\log\mathcal{O}(K, \rho, r)}{K}.
\end{eqnarray}
The following proposition gives the general formula of $\mathcal{E}(\eta)$, under mild technical conditions.

\begin{prop}
\label{prop:general}
Suppose that the following conditions hold:

(i) For every $s \in \mathcal{S}$, the function $J(\gamma, s)/\gamma$ is nonnegative and monotonically non-increasing with $\gamma > 0$, and the limit
\begin{eqnarray}
\dot{J}(0, s) := \lim_{\gamma \downarrow 0} \frac{1}{\gamma}J(\gamma, s)
\end{eqnarray}
exists as an extended number on $\mathbb{R}^+ \cup \{0, \infty\}$;

(ii) The limit $\dot{J}(0, S)$, as a random variable induced by $S$, has its logarithmic moment generating function
\begin{eqnarray}
\Lambda(\lambda) := \log\mathbf{E}\left[\exp(\lambda \dot{J}(0, S_1))\right]
\end{eqnarray}
defined as an essentially smooth and lower semi-continuous function on $\lambda \in \mathbb{R}$.

The wideband outage exponent of the wideband communication system is then given by
\begin{eqnarray}
\mathcal{E}(\eta) = \sup_{\lambda \leq 0} \left\{\frac{\lambda}{\eta} - \Lambda(\lambda)\right\},
\end{eqnarray}
for every $\eta \geq \bar{\eta}$, where
\begin{eqnarray}
\bar{\eta} := \frac{1}{\mathbf{E}[\dot{J}(0, S_1)]}
\end{eqnarray}
is the minimum achievable wideband energy per nat of the given coding scheme.
\end{prop}
{\it Proof:} The proof relies on the theory of large deviations; see, {\it e.g.}, \cite{dembo98:book}. We start with the logarithmic moment generating function of $R(K, \rho)$,
\begin{eqnarray}
\Lambda_K(\lambda) := \log\mathbf{E}\left\{\exp[\lambda R(K, \rho)]\right\}.
\end{eqnarray}
Based upon the sequence $\left\{\Lambda_K(\lambda)\right\}_{K = 1}^\infty$, for each $\lambda \in \mathbb{R}$, we can evaluate the following asymptotic logarithmic moment generating function as
\begin{eqnarray}
\label{eqn:almgf}
\Lambda_\infty(\lambda) &:=& \lim_{K \rightarrow \infty} \frac{1}{K}\Lambda_K(K\lambda)\nonumber\\
&=& \lim_{K \rightarrow \infty} \frac{1}{K} \log\mathbf{E}\left\{\exp[K\lambda R(K, \rho)]\right\}\nonumber\\
&\stackrel{(a)}{=}& \lim_{K \rightarrow \infty} \frac{1}{K} \log\mathbf{E}\left[\prod_{k = 1}^K \exp(K\lambda J(\rho/K, S_k))\right]\nonumber\\
&\stackrel{(b)}{=}& \lim_{K \rightarrow \infty} \frac{1}{K} \log\left\{\mathbf{E}[\exp(K\lambda J(\rho/K, S_1))]\right\}^K\nonumber\\
&=& \lim_{K \rightarrow \infty} \log\mathbf{E}[\exp(K\lambda J(\rho/K, S_1))]\nonumber\\
&\stackrel{(c)}{=}& \log\mathbf{E}\left[\exp(\lambda \cdot\lim_{K\rightarrow \infty}K\cdot J(\rho/K, S_1))\right]\nonumber\\
&\stackrel{(d)}{=}& \log\mathbf{E}[\exp(\lambda \rho \dot{J}(0, S_1))],
\end{eqnarray}
where (a) is from the definition of $R(K, \rho)$, (b) is from the i.i.d. property of $\{S_k\}_{k = 1}^K$, (c) and (d) are from the condition (i). Specifically, (c) is from applying the monotone convergence theorem to the monotonically non-decreasing sequence $\left\{K\cdot J(\rho/K, s)\right\}_{K = 1}^\infty$, and (d) is from the convergence property of $J(\gamma, s)/\gamma$ as $\gamma \downarrow 0$, for every $s \in \mathcal{S}$.

In view of the asymptotic logarithmic moment generating function (\ref{eqn:almgf}), a crucial observation is that $\Lambda_\infty(\lambda)$ can also be obtained as the exact logarithmic moment generating function of the random variable $\rho \dot{J}(0, S_1)$. Consequently, since the G\"{a}rtner-Ellis theorem in conjunction with the condition (ii) ensures the large deviation principle to hold for the sequence $\left\{R(K, \rho)\right\}_{K = 1}^\infty$ (see, {\it e.g.}, \cite[Theorem 2.3.6]{dembo98:book}), this alternative interpretation of $\Lambda_\infty(\lambda)$ asserts that the large deviation rate function and the resulting wideband outage exponent of $\left\{R(K, \rho)\right\}_{K = 1}^\infty$ should be the same as that of the empirical mean sequence $\left\{(\rho/K)\sum_{k = 1}^K \dot{J}(0, S_k)\right\}_{K = 1}^\infty$. Therefore to prove the proposition, it suffices to equivalently characterize the asymptotic behavior of this empirical mean sequence, as follows.

By the law of large numbers, the empirical mean sequence $\left\{(\rho/K)\sum_{k = 1}^K \dot{J}(0, S_k)\right\}_{K = 1}^\infty$ converges in probability to the expectation
\begin{eqnarray}
\bar{r}(\rho) := \rho \mathbf{E}[\dot{J}(0, S_1)].
\end{eqnarray}
Alternatively, the wideband cost per nat converges in probability to
\begin{eqnarray}
\bar{\eta} = \frac{\rho}{\bar{r}(\rho)} = \frac{1}{\mathbf{E}[\dot{J}(0, S_1)]},
\end{eqnarray}
which is the minimum cost per nat that the given transmission scheme can achieve reliably in the wideband limit.

For the empirical mean sequence $\left\{(\rho/K)\sum_{k = 1}^K \dot{J}(0, S_k)\right\}_{K = 1}^\infty$, we can apply Cram\'{e}r's theorem to obtain the large deviation rate function as
\begin{eqnarray}
\Lambda^\ast(x) = \sup_{\lambda \in \mathbb{R}}\{\lambda x - \Lambda_\infty(\lambda)\}.
\end{eqnarray}
For any rate $r \leq \bar{r}(\rho)$, the exponential decay rate
\begin{eqnarray*}
\lim_{K \rightarrow \infty} \frac{-\log \mathrm{Pr}\left[
(\rho/K)\sum_{k = 1}^K \dot{J}(0, S_k) \leq r
\right]}{K},
\end{eqnarray*}
can be further evaluated as \cite[Corollary 2.2.19]{dembo98:book}
\begin{eqnarray}
\lim_{K \rightarrow \infty} \frac{-\log \mathrm{Pr}\left[
(\rho/K)\sum_{k = 1}^K \dot{J}(0, S_k) \leq r
\right]}{K} = \inf_{x \in [0, r]} \Lambda^\ast(x).
\end{eqnarray}
By utilizing the monotonicity property of $\Lambda^\ast(x)$ \cite[Lemma 2.2.5]{dembo98:book}, we have
\begin{eqnarray}
\inf_{x \in [0, r]} \Lambda^\ast(x) = \sup_{\lambda \leq 0}\{\lambda r - \Lambda_\infty(\lambda)\}.
\end{eqnarray}
At this stage, by integrating the established reasoning steps, we have
\begin{eqnarray}
\mathcal{E}(\eta) &=& \lim_{K \rightarrow \infty} \frac{-\log\mathcal{O}(K, \rho, r)}{K}\nonumber\\
&=& \lim_{K \rightarrow \infty} -\frac{1}{K} \log \mathrm{Pr}\left[
(\rho/K)\sum_{k = 1}^K \dot{J}(0, S_k) \leq r
\right]\nonumber\\
&=& \sup_{\lambda \leq 0}\{\lambda r - \Lambda_\infty(\lambda)\}\nonumber\\
&=& \sup_{\lambda \leq 0}\left\{\lambda r - \log\mathbf{E}[\exp(\lambda \rho \dot{J}(0, S_1))]\right\}.
\end{eqnarray}
To conclude the proof, we notice that $\rho$ can be readily absorbed by $\lambda$ via a change of variables, leading to
\begin{eqnarray}
\mathcal{E}(\eta) &=& \sup_{\lambda \leq 0}\left\{\lambda \frac{r}{\rho} - \log\mathbf{E}[\exp(\lambda \dot{J}(0, S_1))]\right\}\nonumber\\
&=& \sup_{\lambda \leq 0} \left\{\frac{\lambda}{\eta} - \Lambda(\lambda)\right\}.
\end{eqnarray}
Thus we establish Proposition \ref{prop:general}.

{\bf Q.E.D.}

In the proof of Proposition \ref{prop:general}, since the achievable rate sequence $\left\{R(K, \rho)\right\}_{K = 1}^\infty$ does not take the form of an empirical mean sequence, we cannot directly invoke Cram\'{e}r's theorem, but instead need to resort to the more general G\"{a}rtner-Ellis theorem to conceptually deduce the large deviation behavior. However, by noticing the asymptotic equivalence of the achievable rate sequence and an induced empirical mean sequence, as bridged by the asymptotic logarithmic moment generating function (\ref{eqn:almgf}), the evaluation of the large deviation behavior is substantially facilitated by utilizing Cram\'{e}r's theorem to the empirical mean sequence. An intuitive way of understanding the proposition is as follows. For sufficiently large $K$, we may approximate the per-channel achievable rate $J(\rho/K, S_k)$ by its first-order expansion term $\dot{J}(0, S_k) \cdot(\rho/K)$, assuming that all the higher-order terms can be safely ignored. Such an approximation immediately leads to
\begin{eqnarray}
R(K, \rho) \approx \frac{\rho}{K} \sum_{k = 1}^K \dot{J}(0, S_k),
\end{eqnarray}
and the wideband outage exponent readily follows from Cram\'{e}r's theorem, as in our proof.

An interesting conclusion of Proposition \ref{prop:general} is that both $\bar{\eta}$ and $\mathcal{E}(\eta)$ depend upon statistics, specifically, the expectation and the logarithmic moment generating function, of only the first-order characteristic of $J(\gamma, s)$ at $\gamma = 0$. Such a property is in contrast to the ergodic case \cite{verdu02:it}, where both first-order and second-order characteristics of the state-averaged rate $\mathbf{E}[J(\gamma, S_1)]$ at $\gamma = 0$ are of relevance. Higher-order characteristics of $J(\gamma, s)$ may well have certain noticeable effect upon the exact value of the outage probability, but they do not affect the wideband outage exponent, which only captures the exponential decay rate of the outage probability.

\section{Scalar Fading}
\label{sec:scalar}

In this section, we apply Proposition \ref{prop:general} to scalar fading channels. For each parallel channel, say, $\mathcal{C}_k$, $k = 1, \ldots, K$, the channel input-output relationship is
\begin{eqnarray}
Y_k = H_k X_k + Z_k,
\end{eqnarray}
where the complex-valued input $X_k$ has an average power constraint $\rho/K$, and the additive white noise $Z_k$ is circular-symmetric complex Gaussian with $Z_k \sim \mathcal{CN}(0, 1)$. The channel state $S_k$ in Section \ref{sec:general} here corresponds to the fading coefficient $H_k$. Let the distribution of $X_k$ be $\mathcal{CN}(0, \rho/K)$, we have\footnote{In fact, more general input distributions, such as symmetric phase-shift keying (PSK) \cite{massey76:isit} and low duty-cycle on-off keying (OOK) even with non-coherent detection \cite{golay49:pire}, can also achieve the same limit function $\dot{J}(0, s)$ as that of the Gaussian inputs; see \cite{verdu02:it} for further elaboration.}
\begin{eqnarray}
\dot{J}(0, H_k) = \lim_{\gamma \downarrow 0} \frac{1}{\gamma} J(\gamma, H_k) = |H_k|^2.
\end{eqnarray}
Therefore Proposition \ref{prop:general} asserts that the wideband minimum energy per nat is $\bar{\eta} = 1/\mathbf{E}[|H_1|^2]$, and for every $\eta \geq \bar{\eta}$, the wideband outage exponent is
\begin{eqnarray}
\label{eqn:ep-scalar}
\mathcal{E}(\eta) = \sup_{\lambda \leq 0}\left\{
\frac{\lambda}{\eta} - \log \mathbf{E}\left[
\exp(\lambda |H_1|^2)
\right]
\right\}.
\end{eqnarray}

In this section, without loss of generality, we normalize the fading coefficient $H_k$ such that $\mathbf{E}[|H_k|^2] = 1$, so that $\bar{\eta} = 1$. We further examine the wideband outage exponent (\ref{eqn:ep-scalar}) by three case studies: Rayleigh, Rician, and Nakagami-$m$.

{\it Rayleigh fading:} The squared fading coefficient $|H_1|^2$ is simply an exponential random variable, therefore the logarithmic moment generating function is $\Lambda(\lambda) = -\log(1 - \lambda)$, and the wideband outage exponent is easily evaluated as
\begin{eqnarray}
\mathcal{E}(\eta) = \frac{1}{\eta} - 1 + \log(\eta),
\end{eqnarray}
for every $\eta \geq 1$.

\begin{figure}[ht]
\psfrag{xlabel}{{$\eta$ (dB)}}
\psfrag{ylabel}{{$\mathcal{E}(\eta)$}}
\epsfxsize=4.5in
\epsfclipon
\centerline{\epsffile{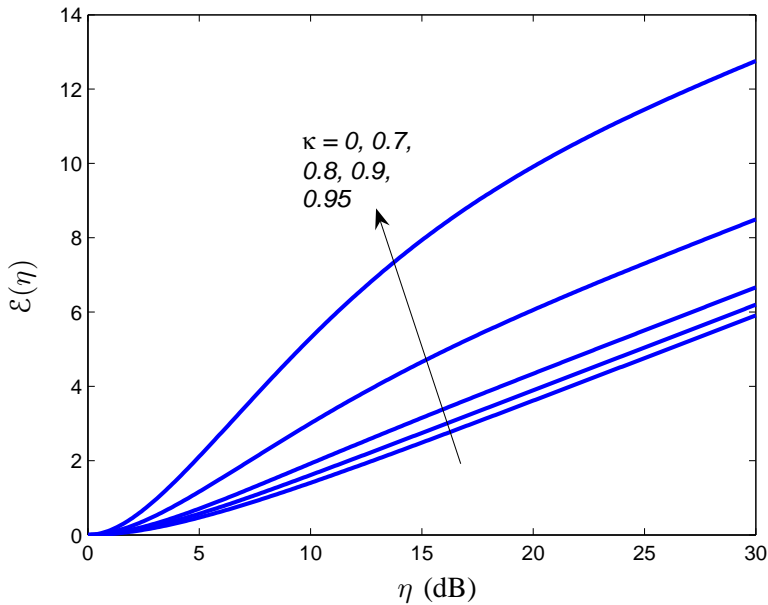}}
\caption{The wideband outage exponent $\mathcal{E}(\eta)$ as a function of the energy per nat $\eta$, for different values of the LOS component $\kappa$ in the Rician fading case. The curve for $\kappa = 0$ corresponds to the Rayleigh fading case.}
\label{fig:rician}
\end{figure}

{\it Rician fading:} We assume that the fading coefficient is $H_1 \sim \mathcal{CN}(\kappa, 1 - \kappa^2)$, where $\kappa \in (0, 1)$ specifies the line-of-sight (LOS) component. Hence $|H_1|^2/(\frac{1 - \kappa^2}{2})$ is a standard noncentral chi-square random variable, with degree of freedom 2 and non-centrality parameter $2\kappa^2/(1 - \kappa^2)$. So the logarithmic moment generating function is
\begin{eqnarray}
\Lambda(\lambda) = \frac{\kappa^2 \lambda}{1 - (1 - \kappa^2)\lambda} - \log\left[1  - (1 - \kappa^2)\lambda\right],
\end{eqnarray}
for $\lambda < \frac{1}{1 - \kappa^2}$. After manipulations it can be evaluated that the wideband outage exponent is
\begin{eqnarray}
\mathcal{E}(\eta) = \frac{1}{(1 - \kappa^2)\eta} + \frac{\kappa^2}{1 - \kappa^2} - \sqrt{1 + \frac{4\kappa^2}{(1 - \kappa^2)^2\eta}} + \log\frac{(1 - \kappa^2)\eta}{2} + \log\left[1 + \sqrt{
1 + \frac{4\kappa^2}{(1 - \kappa^2)^2\eta}
}\right],
\end{eqnarray}
for every $\eta \geq 1$. When $\kappa = 0$ the Rician fading case reduces to the Rayleigh fading case. In Figure \ref{fig:rician} we plot $\mathcal{E}(\eta)$ versus $\eta$ for different values of $\kappa$. Numerically we can observe that $\mathcal{E}(\eta)$ becomes larger as $\kappa$ increases. This is intuitively reasonable, because as $\kappa$ increases the channel becomes the more like a non-faded Gaussian-noise channel. We also observe that, such an increase in $\mathcal{E}(\eta)$ with $\kappa$ is rather abrupt. For $\kappa \leq 0.7$, the gap between the $\mathcal{E}(\eta)$ curves is numerically negligible; but as $\kappa \geq 0.9$, the gap becomes significant.

{\it Nakagami-$m$ fading:} In the Nakagami-$m$ fading case, the squared fading coefficient $|H_1|^2$ is a Gamma distributed random variable with shape parameter $m \geq 1/2$ and scale parameter $1/m$, {\it i.e.}, the probability density function of $|H_1|^2$ is $p(a) = a^{m - 1}\exp(-ma)m^m/\Gamma(m)$ for $a \geq 0$.\footnote{The Rayleigh fading case corresponds to $m = 1$.} The logarithmic moment generating function is therefore $\Lambda(\lambda) = -m\cdot\log(1 - \lambda/m)$, for $\lambda<m$. Consequently, the wideband outage exponent is easily evaluated as
\begin{eqnarray}
\mathcal{E}(\eta) = m\cdot\left[\frac{1}{\eta} - 1 + \log(\eta)\right],
\end{eqnarray}
for every $\eta \geq 1$. Comparing with the Rayleigh fading case, we observe that the wideband outage exponent of the Nakagami-$m$ fading channel is exactly $m$ times that of the Rayleigh fading channel. So for $m > 1$, the requirement on bandwidths of Nakagami-$m$ fading channels is less stringent than that of Rayleigh fading channels. More precisely, in the regime of small outage probabilities, Nakagami-$m$ fading exhibits an approximate $m$-fold savings in bandwidths. For $1/2 \leq m < 1$, Rayleigh fading channels become more bandwidth-efficient, though. Such an $m$-scaling property quantitatively reflects the qualitative intuition that the Nakagami-$m$ fading exhibits less uncertainty as the fading figure $m$ increases.

\section{Multiple Antennas}
\label{sec:mimo}

In this section, we consider channels with multiple inputs and outputs. For each parallel channel, say, $\mathcal{C}_k$, $k = 1, \ldots, K$, the channel equation is
\begin{eqnarray}
\mathbf{Y}_k = \mathbf{H}_k \mathbf{X}_k + \mathbf{Z}_k.
\end{eqnarray}
The complex-valued vector input $\mathbf{X}_k \in \mathbb{C}^{n_\mathrm{t} \times 1}$ has an average power constraint
\begin{eqnarray}
\mathbf{E}[\mathbf{X}_k^\dag \mathbf{X}_k] = \frac{\rho}{K}.
\end{eqnarray}
The temporally i.i.d. additive noise $\mathbf{Z}_k \in \mathbb{C}^{n_\mathrm{r} \times 1}$ is also spatially white, with $\mathbf{Z}_k \sim \mathcal{CN}(\mathbf{0}, \mathbf{I}_{n_\mathrm{r}\times n_\mathrm{r}})$. The fading matrix $\mathbf{H}_k \in \mathbb{C}^{n_\mathrm{r}\times n_\mathrm{t}}$ is characterized by some appropriate probability distribution. For instance, a commonly used fading model is that the elements of $\mathbf{H}_k$ are i.i.d. circular-symmetric complex Gaussian; see, {\it e.g.}, \cite{telatar99:ett}.

Let the input distribution be $\mathbf{X}_k \sim \mathcal{CN}(0, (\rho/K)\mathbf{\Sigma})$, where the covariance matrix $\mathbf{\Sigma} \in \mathbb{R}^{n_\mathrm{t}\times n_\mathrm{t}}$ satisfies $\mathrm{tr}\mathbf{\Sigma} = 1$, where $\mathrm{tr}(\cdot)$ denotes the matrix trace operator. The resulting achievable rate for parallel channel $\mathcal{C}_k$ hence is
\begin{eqnarray}
J(\rho/K, \mathbf{H}_k) = \log\det\left(
\mathbf{I}_{n_\mathrm{r}\times n_\mathrm{r}} + \frac{\rho}{K}\mathbf{H}_k \mathbf{\Sigma} \mathbf{H}_k^\dag
\right),
\end{eqnarray}
and as $K \rightarrow \infty$, we get
\begin{eqnarray}
\label{eqn:jdot-mimo}
\dot{J}(0, \mathbf{H}_k) &=& \frac{1}{\rho} \lim_{K \rightarrow \infty} K\cdot J(\rho/K, \mathbf{H}_k)\nonumber\\
&=& \mathrm{tr}\left[\mathbf{H}_k\mathbf{\Sigma}\mathbf{H}_k^\dag\right].
\end{eqnarray}
From (\ref{eqn:jdot-mimo}) we can then apply Proposition \ref{prop:general} to evaluate the wideband minimum energy per nat and the wideband outage exponent, for different channel models and input covariance structures. We note that, for general MIMO fading channels, characterizing outage-optimal input distributions is still an open problem (see, {\it e.g.}, \cite{telatar99:ett}), therefore the wideband outage exponents obtained in this section should be understood as achievable rather than optimal.

\subsection{Spatially White Inputs}

If we let the input covariance matrix be $\mathbf{\Sigma} = (1/n_\mathrm{t})\cdot \mathbf{I}_{n_\mathrm{t}\times n_\mathrm{t}}$, then we further simplify (\ref{eqn:jdot-mimo}) to
\begin{eqnarray}
\dot{J}(0, \mathbf{H}_k) &=& \frac{1}{n_\mathrm{t}} \mathrm{tr}\left[\mathbf{H}_k\mathbf{H}_k^\dag\right]\nonumber\\
&=& \frac{1}{n_\mathrm{t}} \sum_{i, j} |H_k(i, j)|^2,
\end{eqnarray}
where $H_k(i, j)$ denotes the $i$th-row $j$th-column element of $\mathbf{H}_k$.

{\it Spatially white Rayleigh fading:} The squared fading coefficients $H_k(i, j)$, $i = 1, \ldots, n_\mathrm{r}$, $j = 1, \ldots, n_\mathrm{t}$, are $n_\mathrm{t}n_\mathrm{r}$ i.i.d. exponential random variables, therefore their sum is a chi-square random variable, or, more generally speaking, a Gamma random variable. Consequently, the wideband outage exponent is essentially (except scaling) identical to the scalar Nakagami-$m$ fading case in Section \ref{sec:scalar}, {\it i.e.},
\begin{eqnarray}
\mathcal{E}(\eta) = n_\mathrm{t}n_\mathrm{r} \cdot\left[
\frac{1}{n_\mathrm{r}\eta} - 1 + \log(n_\mathrm{r}\eta)
\right],
\end{eqnarray}
for every $\eta \geq \bar{\eta}:= 1/n_\mathrm{r}$. So besides the $n_\mathrm{r}$-fold decrease in the wideband minimum energy per nat, the multiple antennas further boost the wideband outage exponent by a factor of $n_\mathrm{t}n_\mathrm{r}$, the product of the numbers of transmit and receive antennas, in contrast to the scalar Rayleigh fading case. It is interesting to note that the gain factor $n_\mathrm{t}n_\mathrm{r}$ also appears as the maximum achievable diversity order in the high-SNR asymptotic regime \cite{zheng03:it}.

{\it Spatially correlated Rayleigh fading:} In this case, we assume that each $H_k(i, j)$ follows the marginal distribution $\mathcal{CN}(0, 1)$, and that the vectorization of $\mathbf{H}_k^\dag$, $\mathbf{V}_k := \mathrm{vec}(\mathbf{H}_k^\dag)$, is a correlated circular-symmetric complex Gaussian vector with a covariance matrix $\mathbf{\Psi} \in \mathbb{R}^{n_\mathrm{t}n_\mathrm{r} \times 1}$. It then follows that the moment generating function of $\dot{J}(0, \mathbf{H}_k)$ is
\begin{eqnarray}
\mathbf{E}[\exp(\lambda\dot{J}(0, \mathbf{H}_k))] &=& \int_{\mathbb{C}^{n_\mathrm{t}n_\mathrm{r} \times 1}} \frac{1}{\pi^{n_\mathrm{t}n_\mathrm{r}}\det(\mathbf{\Psi})} \exp\left(-\mathbf{v}^\dag \mathbf{\Psi}^{-1}\mathbf{v}\right)\cdot \exp\left(\frac{ \lambda}{n_\mathrm{t}}\mathbf{v}^\dag \mathbf{v}\right) d\mathbf{v}\nonumber\\
&=& \left[\det\left(\mathbf{I} - \frac{\lambda}{n_\mathrm{t}}\mathbf{\Psi}\right)\right]^{-1},
\end{eqnarray}
for $\lambda < n_\mathrm{t}/\mu_{\max}(\mathbf{\Psi})$, where $\mu_{\max}(\cdot)$ is the maximum eigenvalue of the operand matrix. Consequently, the wideband outage exponent is given by
\begin{eqnarray}
\label{eqn:oe-mimo-corr-white}
\mathcal{E}(\eta) = \sup_{\lambda \leq 0}\left\{
\frac{\lambda}{\eta} + \log\det\left(
\mathbf{I} - \frac{\lambda}{n_\mathrm{t}}\mathbf{\Psi}
\right)
\right\},
\end{eqnarray}
for every $\eta \geq \bar{\eta} = 1/n_\mathrm{r}$.

In order to evaluate (\ref{eqn:oe-mimo-corr-white}), we force the derivative of the function in the bracket to vanish, obtaining the following equation for solving the optimizer $\lambda^\ast$:
\begin{eqnarray}
\sum_{i = 1}^{n_\mathrm{t}n_\mathrm{r}} \frac{\mu_i(\mathbf{\Psi})}{n_\mathrm{t} - \lambda\mu_i(\mathbf{\Psi})} = \frac{1}{\eta},
\end{eqnarray}
where $\mu_i(\cdot)$ is the $i$th eigenvalue of the operand matrix.

\begin{figure}[ht]
\psfrag{xlabel}{{$\eta$ (dB)}}
\psfrag{ylabel}{{$\mathcal{E}(\eta)$}}
\epsfxsize=4.5in
\epsfclipon
\centerline{\epsffile{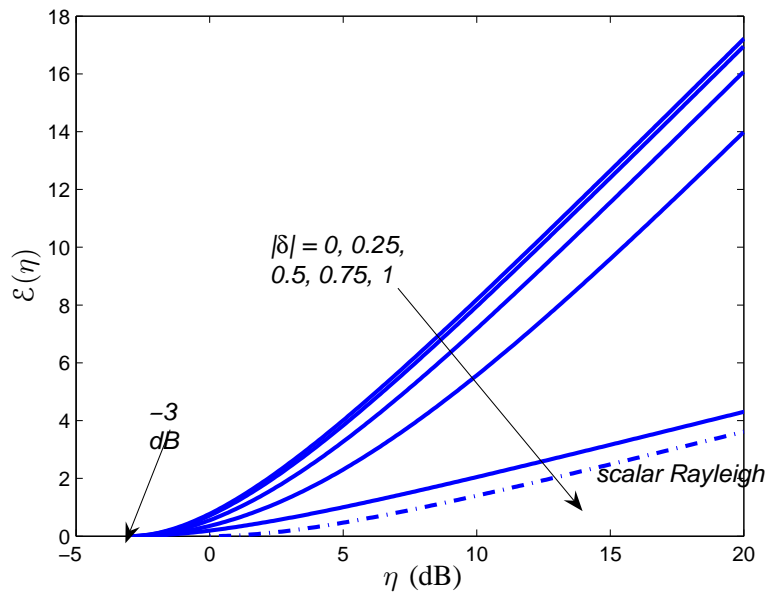}}
\caption{The wideband outage exponent $\mathcal{E}(\eta)$ as a function of the energy per nat $\eta$, for different values of the spatial correlation parameter $|\delta| \leq 1$, under spatially white channel inputs. The dashed-dot curve corresponds to the scalar Rayleigh fading case.}
\label{fig:corr-wi}
\end{figure}

{\it A numerical example:} Consider a two-by-two system with the following covariance matrix of $\mathbf{\Psi}$:
\begin{eqnarray}
\mathbf{\Psi} &=& \left[
\begin{array}{ll}
	1 & \delta\\
	\delta & 1
\end{array}
\right] \otimes \left[
\begin{array}{ll}
	1 & \delta\\
	\delta & 1
\end{array}
\right]\nonumber\\
&=& \left[
\begin{array}{llll}
	1 & \delta & \delta & \delta^2\\
	\delta & 1 & \delta^2 & \delta\\
	\delta & \delta^2 & 1 & \delta\\
	\delta^2 & \delta & \delta & 1
\end{array}
\right],
\end{eqnarray}
where $\otimes$ denotes the Kronecker product, and the parameter $\delta \in [-1, 1]$. As will be seen in the next subsection, this choice of $\mathbf{\Psi}$ is an instance of the separable spatial correlation model \cite{kermoal02:jsac}. Figure \ref{fig:corr-wi} displays $\mathcal{E}(\eta)$ versus $\eta$ for different values of $\delta$, under spatially white channel inputs. Compared to the dashed-dot curve of the scalar Rayleigh fading case, we notice that multiple antennas decrease $\bar{\eta}$ by a factor of $n_\mathrm{r} = 2 \approx 3\mbox{dB}$, regardless of the spatial correlation in $\mathbf{\Psi}$. Comparing the $\mathcal{E}(\eta)$ curves for different value of $\delta$, we notice that as the channel spatial correlation increases ({\it i.e.}, as $|\delta|$ increases), the achieved wideband outage exponent decreases. In the extreme case of full spatial correlation, {\it i.e.}, $|\delta| = 1$, there is no increase in $\mathcal{E}(\eta)$ compared to the scalar Rayleigh case, except for the $3$dB-shift in $\bar{\eta}$.

\subsection{Input Covariance Shaping for Spatially Correlated Fading}

When $\mathbf{\Psi} \neq \mathbf{I}$, there exists spatial correlation among the transmit/receive antennas. Despite its simplicity, the spatially white input $\mathbf{X} \sim \mathcal{CN}\left(\mathbf{0}, \frac{\rho}{n_\mathrm{t}K}\mathbf{I}\right)$ generally even does not optimize the wideband minimum energy per nat among all possible inputs, let alone the wideband outage exponent. As will be shown, for spatially correlated fading channels, the optimal input covariance actually depends upon the target energy efficiency $\eta$.

Utilizing the prior development in this section, we have the following result:
\begin{prop}
\label{prop:oe-corr}
For a spatially correlated fading channel with the fading matrix vectorization $\mathbf{V}_k := \mathrm{vec}(\mathbf{H}_k^\dag)\sim \mathcal{CN}(\mathbf{0}, \mathbf{\Psi})$, suppose that the input covariance matrix is $(\rho/K)\mathbf{\Sigma}$. Then the resulting wideband minimum energy per nat $\bar{\eta}$ and the wideband outage exponent $\mathcal{E}(\eta)$ are respectively given by
\begin{eqnarray}
\label{eqn:rate-corr}
\bar{\eta} &=& \frac{1}{\mathrm{tr}\left((\mathbf{I}_{n_\mathrm{r}\times n_\mathrm{r}} \otimes \mathbf{\Sigma})\mathbf{\Psi}\right)},\nonumber\\
\label{eqn:oe-corr}
\mathcal{E}(\eta) &=& \sup_{\lambda \leq 0} \left\{
\frac{\lambda}{\eta} + \log\det\left(\mathbf{I} - \lambda
(\mathbf{I}_{n_\mathrm{r}\times n_\mathrm{r}} \otimes \mathbf{\Sigma})\mathbf{\Psi}
\right)
\right\}
\end{eqnarray}
for every $\eta \geq \bar{\eta}$, where $\otimes$ denotes the Kronecker product.
\end{prop}
{\it Proof:} We start with evaluating $\dot{J}(0, \mathbf{H}_k)$ from (\ref{eqn:jdot-mimo}). For notational convenience, denote $\mathbf{\Psi}^{1/2} (\mathbf{I}_{n_\mathrm{r}\times n_\mathrm{r}}\otimes \mathbf{\Sigma}) \mathbf{\Psi}^{1/2}$ by $\mathbf{\Phi}$, and its eigenvalues by $\mu_i(\mathbf{\Phi})$, $i = 1, \ldots, n_\mathrm{t}n_\mathrm{r}$. Introduce an auxiliary random vector $\mathbf{V}_{0, k} \sim \mathcal{CN}(\mathbf{0}, \mathbf{I})$ such that we can generate $\mathbf{V}_k \sim \mathcal{CN}(\mathbf{0}, \mathbf{\Psi})$ by $\mathbf{V}_k = \mathbf{\Psi}^{1/2}\mathbf{V}_{0, k}$. We have
\begin{eqnarray}
\dot{J}(0, \mathbf{H}_k) &=& \mathrm{tr}[\mathbf{H}_k \mathbf{\Sigma} \mathbf{H}_k^\dag]\nonumber\\
&=& \sum_{n = 1}^{n_\mathrm{r}} \mathbf{H}_k(n, :) \mathbf{\Sigma} \mathbf{H}_k^\dag(n, :)\nonumber\\
&=& \mathbf{V}_k^\dag (\mathbf{I}_{n_\mathrm{r}\times n_\mathrm{r}}\otimes \mathbf{\Sigma}) \mathbf{V}_k\nonumber\\
&=& \mathbf{V}_{0, k}^\dag \mathbf{\Psi}^{1/2} (\mathbf{I}_{n_\mathrm{r}\times n_\mathrm{r}}\otimes \mathbf{\Sigma}) \mathbf{\Psi}^{1/2} \mathbf{V}_{0, k}.
\end{eqnarray}
By the invariance property under unitary transformation of $\mathcal{CN}(\mathbf{0}, \mathbf{I})$ random vectors, we can further introduce another auxiliary random vector $\tilde{\mathbf{V}}_{0, k} \sim \mathcal{CN}(\mathbf{0}, \mathbf{I})$ such that
\begin{eqnarray}
\label{eqn:aux-proof-corr}
\dot{J}(0, \mathbf{H}_k) &=& \tilde{\mathbf{V}}_{0, k}^\dag \mathrm{diag}(\mu_1(\mathbf{\Phi}), \ldots, \mu_{n_\mathrm{t}n_\mathrm{r}}(\mathbf{\Phi})) \tilde{\mathbf{V}}_{0, k}\nonumber\\
&=& \sum_{i = 1}^{n_\mathrm{t}n_\mathrm{r}} \mu_i(\mathbf{\Phi}) |\tilde{V}_{0, k}(i)|^2.
\end{eqnarray}
Applying Proposition \ref{prop:general} to (\ref{eqn:aux-proof-corr}), we get
\begin{eqnarray}
\bar{\eta} &=& \frac{1}{\mathbf{E}[\dot{J}(0, \mathbf{H}_k)]}\nonumber\\
&=& \left\{\sum_{i = 1}^{n_\mathrm{t}n_\mathrm{r}} \mu_i(\mathbf{\Phi}) \mathbf{E}[|\tilde{V}_{0, k}(i)|^2]\right\}^{-1}\nonumber\\
&=& \left[\sum_{i = 1}^{n_\mathrm{t}n_\mathrm{r}} \mu_i(\mathbf{\Phi})\right]^{-1}\nonumber\\
&=& \frac{1}{\mathrm{tr}[\mathbf{\Phi}]},
\end{eqnarray}
and thus we establish (\ref{eqn:rate-corr}) by utilizing the trace property $\mathrm{tr}[\mathbf{AB}] = \mathrm{tr}[\mathbf{BA}]$.

We proceed to evaluate the logarithmic moment generating function $\Lambda(\lambda)$,
\begin{eqnarray}
\label{eqn:aux-proof-corr-lmgf}
\Lambda(\lambda) &=& \log\mathbf{E}\left[\exp(\lambda \dot{J}(0, \mathbf{H}_k))\right]\nonumber\\
&\stackrel{(a)}{=}& \log \mathbf{E}\left[\prod_{i = 1}^{n_\mathrm{t}n_\mathrm{r}} \exp(\lambda \mu_i(\mathbf{\Phi})|\tilde{V}_{0, k}(i)|^2)\right]\nonumber\\
&\stackrel{(b)}{=}& \log \prod_{i = 1}^{n_\mathrm{t}n_\mathrm{r}} \mathbf{E}\left[\exp(\lambda \mu_i(\mathbf{\Phi})|\tilde{V}_{0, k}(i)|^2)\right]\nonumber\\
&\stackrel{(c)}{=}& -\sum_{i = 1}^{n_\mathrm{t}n_\mathrm{r}} \log(1 - \lambda \mu_i(\mathbf{\Phi}))\nonumber\\
&=& -\log\det(\mathbf{I} - \lambda \mathbf{\Phi})\nonumber\\
&\stackrel{(d)}{=}& -\log\det(\mathbf{I} - \lambda (\mathbf{I}_{n_\mathrm{r}\times n_\mathrm{r}}\otimes \mathbf{\Sigma})\mathbf{\Psi}),
\end{eqnarray}
where (a) is from (\ref{eqn:aux-proof-corr}), (b) is from the mutual independence among the elements of $\tilde{\mathbf{V}}_{0, k}$, (c) is from the fact that $|\tilde{V}_{0, k}(\cdot)|^2$ is an exponential random variable, and (d) is from the determinant property $\det(\mathbf{I} + \mathbf{A}\mathbf{B}) = \det(\mathbf{I} + \mathbf{B}\mathbf{A})$. As we get (\ref{eqn:aux-proof-corr-lmgf}), the wideband outage exponent (\ref{eqn:oe-corr}) readily follows.

{\bf Q.E.D.}

{\it Separable spatial correlation:} A spatial correlation model usually adopted for first-cut analysis is the separable model; see, {\it e.g.}, \cite{kermoal02:jsac}. In this model we can generate $\mathbf{H}_k$ as
\begin{eqnarray}
\mathbf{H}_k = \mathbf{\Psi}_\mathrm{r}^{1/2} \mathbf{H}_{0, k} \mathbf{\Psi}_\mathrm{t}^{1/2},
\end{eqnarray}
where $\mathbf{H}_{0, k} \in \mathbb{C}^{n_\mathrm{r}\times n_\mathrm{t}}$ is a random matrix with i.i.d. $\mathcal{CN}(0, 1)$ elements, and $\mathbf{\Psi}_\mathrm{r}^{1/2}$ ($\mathbf{\Psi}_\mathrm{t}^{1/2}$) is a $n_\mathrm{r} \times n_\mathrm{r}$ ($n_\mathrm{t} \times n_\mathrm{t}$) positive semi-definite deterministic matrix characterizing the spatial correlation at the receive (transmit) side. The overall covariance matrix $\mathbf{\Psi}$ is thus given by
\begin{eqnarray}
\mathbf{\Psi} = \mathbf{\Psi}_\mathrm{r}^{1/2}\otimes \mathbf{\Psi}_\mathrm{t}^{1/2}.
\end{eqnarray}
For the separable spatial correlation model, we can further reduce the expressions of $\bar{\eta}$ and $\mathcal{E}(\eta)$ in Proposition \ref{prop:oe-corr} to
\begin{eqnarray}
\bar{\eta} &=& \frac{1}{\mathrm{tr}(\mathbf{\Psi}_\mathrm{r}) \mathrm{tr}(\mathbf{\Sigma}\mathbf{\Psi}_\mathrm{t})},\nonumber\\
\mathcal{E}(\eta) &=& \sup_{\lambda \leq 0} \left\{\frac{\lambda}{\eta} + \sum_{i = 1}^{n_\mathrm{t}}\sum_{j = 1}^{n_\mathrm{r}}\log(1 - \lambda \mu_i(\mathbf{\Sigma}\mathbf{\Psi}_\mathrm{t})\mu_j(\mathbf{\Psi}_\mathrm{r}))\right\}.
\end{eqnarray}

{\it A covariance-shaping problem:} In light of Proposition \ref{prop:oe-corr}, we can formulate the following input covariance shaping problem to maximize the resulting wideband outage exponent for a given target energy per nat $\eta$:
\begin{eqnarray*}
\max_{\mathbf{\Sigma}} \mathcal{E}(\eta), \quad\mbox{s.t.}\; \bar{\eta} \leq \eta, \mathbf{\Sigma}\;\mbox{is positive semi-definite, and}\;\mathrm{tr}\mathbf{\Sigma} = 1.
\end{eqnarray*}
Here we note that $\bar{\eta}$ also depends upon the choice of $\mathbf{\Sigma}$. Although its log-determinant part is concave in $\mathbf{\Sigma}$, the function $\mathcal{E}(\eta)$ itself is generally not concave due to its supremum operation. Therefore the covariance-shaping problem may only be numerically computed for general spatial correlation $\mathbf{\Psi}$.

{\it An illustrative example:} We examine a simple example to conclude this section. Consider a two-input-single-output system, with the fading vector $\mathbf{H}_k = [H_k(1) H_k(2)]$ of the $k$th parallel channel having cross-correlation $0 \leq \delta < 1$, {\it i.e.},
\begin{eqnarray*}
\mathbf{H}_k^\dag \sim \mathcal{CN}\left(0, \mathbf{\Psi} = \left[
\begin{array}{ll}
	1 & \delta\\
	\delta & 1
\end{array}
\right]\right).
\end{eqnarray*}
Without loss of generality, the input covariance matrix is restricted to be statistically symmetric between the two transmit antennas, and is thus parametrized by
\begin{eqnarray*}
\mathbf{\Sigma} = \frac{1}{2}\left[
\begin{array}{ll}
	1 & \xi\\
	\xi & 1
\end{array}
\right],
\end{eqnarray*}
where $-1 \leq \xi \leq 1$ satisfies the positive semi-definite constraint.

The wideband minimum energy per nat $\bar{\eta}$, as a function of $\xi$, is $\bar{\eta}(\xi) = (1 + \delta \xi)^{-1}$. So the optimal $\xi$ in terms of minimizing $\bar{\eta}$ is $\xi = 1$, {\it i.e.}, fully correlated antennas, achieving $\bar{\eta}(1) = (1 + \delta)^{-1}$. In contrast, for uncorrelated antennas with $\xi = 0$, we have $\bar{\eta}(0) = 1$. Thus as the target $\eta$ decreases, it eventually becomes optimal to employ fully correlated antennas. But, if the target $\eta$ is large, then it can be more preferable to adjust $\xi$ such that the resulting $\mathcal{E}(\eta)$ outperforms that of $\xi = 1$. To see this, we evaluate the asymptotic behavior of $\mathcal{E}(\eta)$ as $\eta \rightarrow \infty$, and after some manipulations, we find
\begin{eqnarray}
\mathcal{E}(\eta) = \left\{
\begin{array}[pos]{ll}
	\log \eta + \log(1 - \delta) - 1 + o(1) & \mbox{if}\; \xi = -1\\
	2 \log \eta + \log(1 - \delta^2) + \log(1 - \xi^2) - 2 + o(1) & \mbox{if}\; -1 < \xi < 1\\
	\log \eta + \log(1 + \delta) - 1 + o(1) & \mbox{if}\; \xi = 1
\end{array}
\right.
\end{eqnarray}
as $\eta \rightarrow \infty$. Thus as the target $\eta$ increases, it eventually becomes optimal to employ uncorrelated antennas ($\xi = 0$).

\begin{figure}[ht]
\psfrag{xlabel}{{$\eta$ (dB)}}
\psfrag{ylabel}{{$\mathcal{E}(\eta)$}}
\epsfxsize=4.5in
\epsfclipon
\centerline{\epsffile{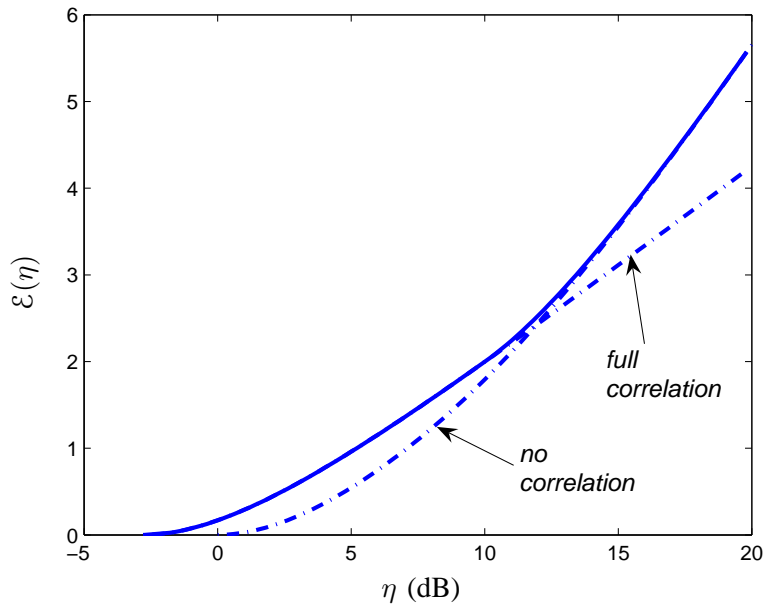}}
\caption{The wideband outage exponent $\mathcal{E}(\eta)$ as a function of the energy per nat $\eta$, optimized over the input cross-correlation $\xi$. The channel spatial correlation parameter is taken as $\delta = 0.9$. The dashed-dot curves correspond to fully correlated inputs and uncorrelated inputs, as indicated respectively.}
\label{fig:corr-shap}
\end{figure}

Figure \ref{fig:corr-shap} displays the optimized $\mathcal{E}(\eta)$ versus $\eta$ with $\delta = 0.9$. The dashed-dot curves are the $\mathcal{E}(\eta)$ curves for $\xi = 0$ and $1$, respectively. It is evidently shown that while fully correlated inputs ($\xi = 1$) attain optimality for small $\eta$, as $\eta$ gets large uncorrelated inputs ($\xi = 0$) become optimal.

\section{One-Bit Channel State Feedback}
\label{sec:feedback}

Since fading channels exhibit abundant diversity in various forms, even a limited amount of channel state information via feedback enables the transmitter to adapt its power and rate allocation such that the resulting performance can be dramatically improved; see, {\it e.g.}, \cite{love04:commag} and references therein. The benefit becomes the most significant for wideband systems. It has been shown that \cite{sun03:isit}, wideband systems with merely one-bit (per parallel channel) CSI at the transmitter achieve essentially the same performance as wideband systems with full transmit CSI, in the sense that their achievable information rates can both be made arbitrarily large with their ratio asymptotically approaching one. Even though such a divergent property of information rates heavily depends upon the underlying channel fading model with an infinite support set (see \cite[Sec. II.A]{stein87:jsac} for some comments on this issue), it is still a reasonable expectation that the asymptotic results, if appropriately interpreted, will provide useful insights into real-world practice.

In this section, we proceed to investigate the effect of limited transmit CSI, from the perspective of the wideband outage exponent. For concreteness, we analyze a specific two-level power allocation scheme with one-bit channel state feedback, for the scalar Rayleigh fading channel model. It is conceptually straightforward to formulate similar problems for systems with finer power allocation levels and channel state quantization levels, or with more general fading models. However, there the analytical approach quickly turns out to be intractable and unilluminating, and instead a heavy dose of numerical recipe becomes necessary, which is beyond the scope of this paper.

\subsection{Transmission Protocol and Wideband Minimum Energy per Nat}
\label{subsec:protocol}

Consider the Rayleigh fading model as in Section \ref{sec:scalar}. For each parallel channel, the receiver informs the transmitter an one-bit feedback
\begin{eqnarray}
F_k = \left\{
\begin{array}{ll}
	0 & \mbox{if}\;|H_k|^2 \leq \tau\\
	1 & \mbox{otherwise}
\end{array}
\right.,
\end{eqnarray}
where $\tau > 0$ is a protocol parameter. Since $|H_k|^2$ is an exponential random variable, the probabilities of $F_k$ are $p_0 := \mathrm{Pr}[F_k = 0] = 1 - \exp(-\tau)$ and $p_1 := \mathrm{Pr}[F_k = 1] = \exp(-\tau)$. Denote the number of parallel channels with $F_\cdot = 0$ by $K_0$, which is a binomial random variable, or, the sum of $K$ i.i.d. Bernoulli random variables. Denote $(K - K_0)$ by $K_1$, which is also binomial. For each parallel channel with $F_\cdot = 0$, we allocate an input power $(g_0/K_0)\rho$; and for each with $F_\cdot = 1$, we allocate an input power $(g_1/K_1)\rho$. Here the protocol parameters $g_0 \in [0, 1]$ and $g_1 = 1 - g_0$ characterize the power allocation scheme.\footnote{If $K_0 = 0$ or $K_1 = 0$, then the ``reserved'' power $g_0 \rho$ or $g_1 \rho$ is not actually utilized and thus wasted. In this paper, we do not allow the transmitter to adaptively adjust $g_0$ according to the realization of $K_0$, for analytical simplicity.} By noting that the $K$ parallel channels are mutually independent, we may write the total achievable rate as
\begin{eqnarray}
\label{eqn:rate-raw}
R(K, \rho) = \left\{
\begin{array}{ll}
	\sum_{j = 1}^K \log\left(1 + \frac{g_1\rho}{K} |\tilde{H}^{(1)}_j|^2\right) & \mbox{if}\;K_0 = 0\\
	\sum_{i = 1}^{K_0} \log\left(1 + \frac{g_0\rho}{K_0} |\tilde{H}^{(0)}_i|^2\right) + \sum_{j = 1}^{K_1} \log\left(1 + \frac{g_1\rho}{K_1} |\tilde{H}^{(1)}_j|^2\right) & \mbox{if}\;K_0 \neq 0\;\mbox{or}\;K\\
	\sum_{i = 1}^K \log\left(1 + \frac{g_0\rho}{K} |\tilde{H}^{(0)}_i|^2\right) & \mbox{if}\;K_0 = K
\end{array}
\right.
\end{eqnarray}
where, $K_0$ is a binomial random variable with
\begin{eqnarray}
\mathrm{Pr}[K_0 = k] = C_K^k p_0^k p_1^{K - k},\quad\mbox{for}\;k = 0, 1, \ldots, K;
\end{eqnarray}
$|\tilde{H}^{(0)}_i|^2$, $i = 1, \ldots, K_0$, are samples from a sequence of i.i.d. random variables with a (truncated exponential) density function $f_0(x) = \exp(-x)/p_0$, for $x \in [0, \tau]$; and similarly, $|\tilde{H}^{(1)}_j|^2$, $j = 1, \ldots, K_1 = K - K_0$, are samples from a sequence of i.i.d. random variables with $f_1(x) = \exp(-x)/p_1$, for $x \in (\tau, \infty)$. In the sequel, for notational convenience, we denote $|\tilde{H}^{(0)}_i|^2$ by $A^{(0)}_i$, and $|\tilde{H}^{(1)}_j|^2$ by $A^{(1)}_j$.

As $K \rightarrow \infty$, the total rate $R(K, \rho)$ converges in probability to a deterministic quantity, which in turn gives the wideband minimum energy per nat, as established by the following proposition.
\begin{prop}
\label{prop:rate-fdbk}
For a given transmission protocol with one-bit channel state feedback, as $K \rightarrow \infty$, it achieves the wideband minimum energy per nat
\begin{eqnarray}
\bar{\eta} = \left[\tau + 1 - \frac{g_0 \tau}{1 - \exp(-\tau)}\right]^{-1}.
\end{eqnarray}
\end{prop}
{\it Proof:} Since the binomial random variable $K_0$ is the sum of $K$ i.i.d. Bernoulli random variables, as $K \rightarrow \infty$, from the law of large numbers, we have for any arbitrarily small $\epsilon > 0$,
\begin{eqnarray}
\mathrm{Pr}\left[\frac{K_0}{K} \in [p_0 - \epsilon, p_0 + \epsilon]\right] \rightarrow 1.
\end{eqnarray}
Thus we only need to consider the case in which the ``typical'' event $\left\{\frac{K_0}{K} \in [p_0 - \epsilon, p_0 + \epsilon]\right\}$ is true. In view of the total achievable rate (\ref{eqn:rate-raw}), we notice that, for every set of realizations of the $K$ parallel channels in this typical event, $R(K, \rho)$ can be bounded as
\begin{eqnarray}
\label{eqn:rate-lb}
R(K, \rho) \geq \sum_{i = 1}^{(p_0 - \epsilon)K} \log\left(1 + \frac{g_0 \rho}{(p_0 + \epsilon)K} A^{(0)}_i\right) + \sum_{j = 1}^{(p_1 - \epsilon)K} \log\left(1 + \frac{g_1 \rho}{(p_1 + \epsilon)K} A^{(1)}_j\right),\\
\label{eqn:rate-ub}
R(K, \rho) \leq \sum_{i = 1}^{(p_0 + \epsilon)K} \log\left(1 + \frac{g_0 \rho}{(p_0 - \epsilon)K} A^{(0)}_i\right) + \sum_{j = 1}^{(p_1 + \epsilon)K} \log\left(1 + \frac{g_1 \rho}{(p_1 - \epsilon)K} A^{(1)}_j\right).
\end{eqnarray}
Here we implicitly assume that $(p_0 \pm \epsilon)K$ are integer-valued, a restriction becoming immaterial as $K \rightarrow \infty$. For each of the four summations in (\ref{eqn:rate-lb}) and (\ref{eqn:rate-ub}), we can invoke Proposition \ref{prop:general} to establish its convergence in probability as $K \rightarrow \infty$. For example,
\begin{eqnarray}
\sum_{i = 1}^{(p_0 - \epsilon)K} \log\left(1 + \frac{g_0 \rho}{(p_0 + \epsilon)K} A^{(0)}_i\right) &=& \sum_{i = 1}^{K^\prime} \log\left(1 + \frac{(p_0 - \epsilon)g_0 \rho}{(p_0 + \epsilon)K^\prime} A^{(0)}_i \right) \;\;[\mbox{let}\;K^\prime = (p_0 - \epsilon)K]\nonumber\\
&\rightarrow& \rho \mathbf{E}\left[\lim_{\gamma \downarrow 0} \frac{1}{\gamma} \log\left(
1 + \frac{(p_0 - \epsilon)g_0 \gamma}{p_0 + \epsilon} A^{(0)}_1
\right)\right]\nonumber\\
&=& g_0 \rho \frac{p_0 - \epsilon}{p_0 + \epsilon} \mathbf{E}[A^{(0)}_1],\quad\;\;\mbox{in probability}.
\end{eqnarray}
Following the similar procedure for the other three summations, we obtain
\begin{eqnarray}
\label{eqn:rate-ineq}
g_0 \rho \frac{p_0 - \epsilon}{p_0 + \epsilon} \mathbf{E}[A^{(0)}_1] + g_1 \rho \frac{p_1 - \epsilon}{p_1 + \epsilon} \mathbf{E}[A^{(1)}_1] \leq R(K, \rho) \leq g_0 \rho \frac{p_0 + \epsilon}{p_0 - \epsilon} \mathbf{E}[A^{(0)}_1] + g_1 \rho \frac{p_1 + \epsilon}{p_1 - \epsilon} \mathbf{E}[A^{(1)}_1],
\end{eqnarray}
in probability as $K \rightarrow \infty$. Since we can choose $\epsilon > 0$ as an arbitrarily small positive number, the two sides of (\ref{eqn:rate-ineq}) can be made arbitrarily close, and therefore we have
\begin{eqnarray}
\lim_{K \rightarrow \infty} R(K, \rho) &=& g_0 \rho \mathbf{E}[A^{(0)}_1] + g_1 \rho \mathbf{E}[A^{(1)}_1]\nonumber\\
&=& \frac{g_0 \rho}{p_0}\int_0^\tau e^{-x} x dx + \frac{g_1 \rho}{p_1} \int_\tau^\infty e^{-x} x dx\nonumber\\
&=& \left[\tau + 1 - \frac{g_0 \tau}{1 - \exp(-\tau)}\right] \rho, \quad\mbox{in probability},
\end{eqnarray}
which immediately yields the wideband minimum energy per nat $\bar{\eta}$.

{\bf Q.E.D.}

\begin{figure}[ht]
\psfrag{xlabel}{{$\tau$ (dB)}}
\psfrag{ylabel}{{$\bar{\eta}$}}
\epsfxsize=4.5in
\epsfclipon
\centerline{\epsffile{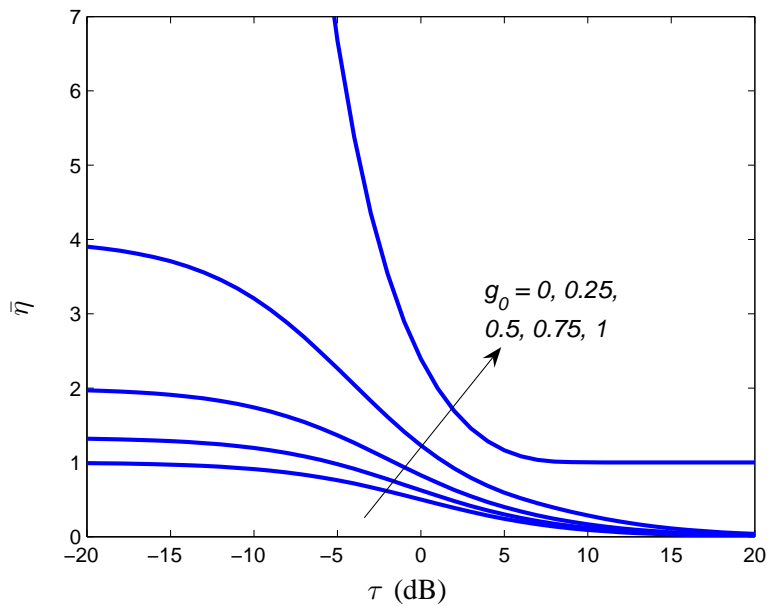}}
\caption{The wideband minimum energy per nat $\bar{\eta}$ as a function of $\tau$ for different values of $g_0$.}
\label{fig:ratefdbk}
\end{figure}

From Proposition \ref{prop:rate-fdbk} we observe that $\bar{\eta}$ can be made arbitrarily close to zero by increasing $\tau$. For small and large values of $\tau$, we have respectively
\begin{eqnarray}
\bar{\eta} &\sim& \frac{1}{1 - g_0},\quad\quad\tau \ll 1;\\
\bar{\eta} &\sim& \frac{1}{(1 - g_0)\tau + 1},\quad\quad\tau \gg 1.
\end{eqnarray}
Figure \ref{fig:ratefdbk} displays $\bar{\eta}$ as a function of $\tau$ for different values of $g_0$.

\subsection{Wideband Outage Exponent: On-Off Power Allocation}
\label{subsec:oe-onoff}

From Proposition \ref{prop:rate-fdbk}, it appears always beneficial to increase $\tau$, because this leads to small $\bar{\eta}$ which implies improved energy efficiency in the wideband limit. However, this perspective overlooks the outage probability. Indeed, as will be shown shortly, increasing $\tau$ generally has a rather detrimental impact upon the wideband outage exponent. Therefore a tradeoff exists between $\bar{\eta}$ and $\mathcal{E}(\eta)$, leveraged by the protocol parameters $\tau$ and $g_0$. In this subsection, we single out the special case of on-off power allocation with $g_0 = 0$, {\it i.e.}, the transmitter only uses those parallel channels with squared fading coefficient greater than $\tau$. In the next subsection, based upon the key insights obtained from the on-off power allocation case, we will further establish general results for $0 \leq g_0 \leq 1$.

Specifying $g_0 = 0$ in Proposition \ref{prop:rate-fdbk}, we get $\bar{\eta} = 1/(\tau + 1)$, in order words, the achievable rate converges to $(\tau + 1)\rho$ in the wideband limit. To evaluate $\mathcal{E}(\eta)$, it is useful to first perform the following qualitative reasoning. Roughly speaking, as long as there are sufficiently many parallel channels with $F_\cdot = 1$, then the achievable rate will be bound to exceed $\tau \rho$, because the on-off power allocation effectively converts the original wideband Rayleigh fading channel into a wideband fading channel with squared fading coefficients always larger than $\tau$. So the ``threshold rate'' $\tau\rho$, or equivalently, the ``threshold $\eta$'' $1/\tau$, has a special role in the on-off power allocation scheme. For rates below this threshold rate, the dominant outage event occurs when the number of ``on'' parallel channels, $K_1$, does not grow to infinity with $K$. Following this line of thought, we can establish the following proposition characterizing the wideband outage exponent.
\begin{prop}
\label{prop:oe-onoff}
For a given transmission protocol with one-bit channel state feedback and on-off power allocation, the wideband outage exponent is given by
\begin{eqnarray}
\label{eqn:oe-onoff}
\mathcal{E}(\eta) = \left\{
\begin{array}{ll}
	\tau + (1 - x^\ast)\left[{1}/{\eta} - \tau - 1 - \log\left({1}/{\eta} - \tau\right)\right] - x^\ast \log(e^\tau - 1) - H_\mathrm{b}(x^\ast) & \\
	& \hspace{-1in}\mbox{if}\; {1}/{(\tau + 1)} \leq \eta \leq {1}/{\tau}\\
	\tau - \log(e^\tau - 1) & \hspace{-2in}\mbox{if}\; \eta > {1}/{\tau}
\end{array}
\right.
\end{eqnarray}
where
\begin{eqnarray}
x^\ast = \frac{(e^\tau - 1)\exp(1/\eta - \tau - 1)}{(e^\tau - 1)\exp(1/\eta - \tau - 1) + 1/\eta - \tau},
\end{eqnarray}
and $H_\mathrm{b}(x) := -x\log x - (1 - x)\log(1 - x)$ is the binary entropy function.
\end{prop}
{\it Proof:} We prove this proposition as a special case in the general proof of Proposition \ref{prop:oe-general}. {\bf Q.E.D.}

The wideband outage exponent $\mathcal{E}(\eta)$ is monotonically non-decreasing with $\eta$. For small and large values of $\tau$, the large-$\eta$ portion of $\mathcal{E}(\eta)$, $\tau - \log(e^\tau - 1)$, exhibits the following behavior:
\begin{eqnarray}
\tau - \log(e^\tau - 1) &\sim& \log(1/\tau),\quad\quad\tau \ll 1;\\
\tau - \log(e^\tau - 1) &\sim& \exp(-\tau),\quad\quad\tau \gg 1.
\end{eqnarray}

\begin{figure}[ht]
\psfrag{xlabel}{{$\eta$} (dB)}
\psfrag{ylabel}{{$\mathcal{E}(\eta)$}}
\epsfxsize=4.5in
\epsfclipon
\centerline{\epsffile{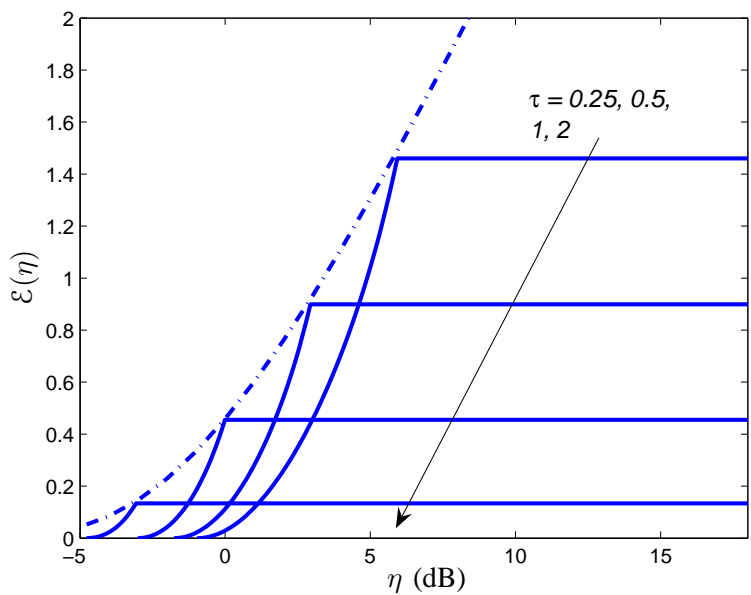}}
\caption{The wideband outage exponent $\mathcal{E}(\eta)$ as a function of the energy per nat $\eta$ for the on-off power allocation scheme, for different values of $\tau$. The dashed-dot curve is the upper envelope of the $\mathcal{E}(\eta)$ over all $\tau$.}
\label{fig:fdbk-onoff}
\end{figure}

For small $\tau$, the wideband outage exponent slowly grows large for $\eta > 1/\tau$. This has been observed for the feedback-less case in Section \ref{sec:scalar}, which may be roughly viewed as the limiting case for $\tau \rightarrow 0$. For large $\tau$, the wideband outage exponent rapidly (at an exponential speed) decreases to zero. Such a decrease may be interpreted as the penalty for the benefit in $\bar{\eta}$. Figure \ref{fig:fdbk-onoff} displays curves of $\mathcal{E}(\eta)$ as a function of $\eta$, for $\tau = 0.25, 0.5, 1, 2$. The tradeoff between $\bar{\eta}$ and $\mathcal{E}(\eta)$ is evidently visible.

For a given target $\eta$, it is then important to adjust $\tau$ to maximize the resulting $\mathcal{E}(\eta)$. This maximum wideband outage exponent can be obtained by plotting all the $\mathcal{E}(\eta)$ versus $\eta$ curves for different values of $\tau$, then taking their upper envelop. Analytically, from (\ref{eqn:oe-onoff}) we can find that the upper envelope exactly corresponds to those turning points $\left(1/\tau, \tau - \log(e^\tau - 1)\right)$ in the first quadrant, and Figure \ref{fig:fdbk-onoff} also displays this upper envelope in dashed-dot curve. For example, in order to achieve $\eta = 1 = 0 \mbox{dB}$, from the dashed-dot curve we find that the on-off power allocation scheme should have $\tau = 1$, yielding the maximum wideband outage exponent $\mathcal{E}(1) \approx 0.45$.

\subsection{Wideband Outage Exponent: General Case}

For general transmission protocols with $g_0 > 0$, we can analogously identify a threshold rate as in the on-off power allocation case. Following similar reasoning as in Section \ref{subsec:oe-onoff}, we can find that the threshold rate is in general given by $r_c := (1 - g_0)\tau\rho$. Below this threshold rate, the dominant outage event occurs when $K_1$ does not grow to infinity with $K$. The following proposition gives the wideband outage exponent for the general case.
\begin{prop}
\label{prop:oe-general}
For a given transmission protocol with one-bit channel state feedback, the wideband outage exponent is given by
\begin{eqnarray}
\mathcal{E}(\eta) = \left\{
\begin{array}{ll}
  \min\left\{ \inf_{x \in (0, 1)} \tilde{\mathcal{E}}(\eta, x), \tilde{\mathcal{E}}_0(\eta)\right\} & \mbox{if}\;\bar{\eta} \leq \eta \leq \frac{1}{(1 - g_0)\tau}\\
	\tilde{\mathcal{E}}_0(\eta) & \mbox{if}\; \eta > \frac{1}{(1 - g_0)\tau}
\end{array}
\right.,
\end{eqnarray}
where
\begin{eqnarray}
\tilde{\mathcal{E}}_0(\eta) &:=& \sup_{\lambda \leq 0}\left\{\frac{\lambda}{\eta} + \log(1 - g_0 \lambda) - \log\left[1 - e^{-(1 - g_0 \lambda)\tau}\right]\right\},\\
\tilde{\mathcal{E}}(\eta, x) &:=& \sup_{\lambda \leq 0}\left\{\frac{\lambda}{\eta} - x \log\left[e^{(1 - g_0 \lambda/x)\tau} - 1\right] + x\log(x - g_0 \lambda)\right.\nonumber\\ &&\quad\quad \Bigl.+ (1 - x)\log[1 - x - (1 - g_0)\lambda] + (1 - \lambda)\tau\Bigl\}.
\end{eqnarray}
\end{prop}
{\it Proof:} In light of the intuitive reasoning on the threshold rate $r_c$, we establish the wideband outage exponent for two cases: rates below $r_c$ and rates above $r_c$, separately.

Case [a] ($r < r_c$):

Conditioning upon $K_0$ and fixing an arbitrarily small $\epsilon > 0$, we can expand the outage probability as
\begin{eqnarray}
\label{eqn:general-case-a}
\mathcal{O}(K, \rho, r) &=& \mathrm{Pr}[R(K, \rho) \leq r]\nonumber\\
&=& \mathrm{Pr}\left[R(K, \rho) \leq r\left|\frac{K_0}{K} < 1 - \epsilon\right.\right]\cdot\mathrm{Pr}\left[\frac{K_0}{K} < 1 - \epsilon\right] \nonumber\\
&&\quad\quad + \mathrm{Pr}\left[R(K, \rho) \leq r\left|\frac{K_0}{K} \geq 1 - \epsilon\right.\right]\cdot\mathrm{Pr}\left[\frac{K_0}{K} \geq 1 - \epsilon\right].
\end{eqnarray}

For sufficiently large $K$, the first sum in (\ref{eqn:general-case-a}) becomes strictly zero, as justified by the following lemma.
\begin{lem}
\label{lem:general-case-a}
For any fixed $r < r_c$ and $\epsilon > 0$, the probability $\mathrm{Pr}\left[R(K, \rho) \leq r\left|\frac{K_0}{K} < 1 - \epsilon\right.\right]\cdot\mathrm{Pr}\left[\frac{K_0}{K} < 1 - \epsilon\right]$ is equal to zero for every sufficiently large $K$.
\end{lem}
{\it Proof of Lemma \ref{lem:general-case-a}:} The condition $K_0/K < 1 - \epsilon$ implies that the number of parallel channels with $F_\cdot = 1$, {\it i.e.}, with squared fading coefficients larger than $\tau$, grows to infinity with $K$. For any realization of $K_0 = k_0$ satisfying $k_0 < (1 - \epsilon)K$, the sum rate $R(K, \rho)$ is always larger than the deterministic quantity
\begin{eqnarray*}
(K - k_0)\log\left(1 + \frac{(1 - g_0)\tau\rho}{K - k_0}\right),
\end{eqnarray*}
which uniformly converges to $r_c = (1 - g_0)\tau\rho$ as $K \rightarrow \infty$. Thus Lemma \ref{lem:general-case-a} follows.

Based upon Lemma \ref{lem:general-case-a}, for sufficiently large $K$ we have
\begin{eqnarray}
\label{eqn:general-case-a-1}
\mathcal{O}(K, \rho, r) = \mathrm{Pr}\left[R(K, \rho) \leq r\left|\frac{K_0}{K} \geq 1 - \epsilon\right.\right]\cdot\mathrm{Pr}\left[\frac{K_0}{K} \geq 1 - \epsilon\right].
\end{eqnarray}
To upper bound $\mathcal{O}(K, \rho, r)$, we note that the first term in (\ref{eqn:general-case-a-1}) can be upper bounded as
\begin{eqnarray*}
\mathrm{Pr}\left[R(K, \rho) \leq r\left|\frac{K_0}{K} \geq 1 - \epsilon\right.\right] \leq \mathrm{Pr}[\tilde{R}(K, \rho) \leq r],
\end{eqnarray*}
where
\begin{eqnarray*}
\tilde{R}(K, \rho) := \sum_{i = 1}^{(1 - \epsilon)K} \log\left(1 + \frac{g_0\rho}{K} A_i^{(0)}\right)
\end{eqnarray*}
is always no greater than $R(K, \rho)$ because in the sum we curtail all the terms with index greater than $(1 - \epsilon)K$.

To lower bound $\mathcal{O}(K, \rho, r)$, we rewrite (\ref{eqn:general-case-a-1}) as
\begin{eqnarray}
\mathcal{O}(K, \rho, r) &=& \mathrm{Pr}\left[R(K, \rho) \leq r\left|\frac{K_0}{K} \geq 1 - \epsilon\right.\right]\cdot\mathrm{Pr}\left[\frac{K_0}{K} \geq 1 - \epsilon\right]\nonumber\\
&=& \mathrm{Pr}\left[R(K, \rho) \leq r, \frac{K_0}{K} \geq 1 - \epsilon\right]\nonumber\\
&\geq& \mathrm{Pr}\left[R(K, \rho) \leq r, K_0 = K\right]\nonumber\\
&=& \mathrm{Pr}\left[R(K, \rho) \leq r| K_0 = K\right]\cdot \mathrm{Pr}[K_0 = K].
\end{eqnarray}

From the upper and lower bounds, by utilizing the theory of large deviations as in Proposition \ref{prop:general}, we can bound the exponential decay rate of $\mathcal{O}(K, \rho, r)$, {\it i.e.}, the wideband outage exponent, as
\begin{eqnarray}
\label{eqn:general-case-a-lb}
\mathcal{E}(\eta) &\geq& \sup_{\lambda \leq 0} \left\{
\frac{\lambda}{\eta} - \log \mathbf{E}\left[\exp\left((1 - \epsilon)g_0 \lambda A_1^{(0)}\right)\right]
\right\} \nonumber\\
&& \quad\quad + \inf_{x\in [1 - \epsilon, 1]}\left\{x \log\frac{p_1}{p_0} - \log p_1 - H_\mathrm{b}(x)\right\}\\
\label{eqn:general-case-a-ub}
\mathcal{E}(\eta) &\leq& \sup_{\lambda \leq 0} \left\{
\frac{\lambda}{\eta} - \log \mathbf{E}\left[\exp\left(g_0 \lambda A_1^{(0)}\right)\right]
\right\} - \log p_0.
\end{eqnarray}
Because the number $\epsilon > 0$ can be made arbitrarily small, as $\epsilon \downarrow 0$ the lower bound (\ref{eqn:general-case-a-lb}) coincides with the upper bound (\ref{eqn:general-case-a-ub}), and the value of $\mathcal{E}(\eta)$ is thus determined as
\begin{eqnarray}
\mathcal{E}(\eta) &=& \sup_{\lambda \leq 0} \left\{\frac{\lambda}{\eta} - \log \mathbf{E}\left[\exp(g_0 \lambda A_1^{(0)})\right]\right\} - \log p_0\nonumber\\
&=& \sup_{\lambda \leq 0} \left\{\frac{\lambda}{\eta} - \log \left[\frac{1}{p_0}\int_0^\tau e^{g_0 \lambda x} e^{-x} dx \right]\right\} - \log p_0\nonumber\\
&=& \tilde{\mathcal{E}}_0(\eta) := \sup_{\lambda \leq 0} \left\{\frac{\lambda}{\eta} + \log(1 - g_0 \lambda) - \log\left[1 - e^{-(1 - g_0 \lambda)\tau}\right]\right\}.
\end{eqnarray}
The preceding derivation of $\mathcal{E}_0(\eta)$ applies to general transmission protocols. For the special case of the on-off power allocation, it is not necessary to invoke the theory of large deviations. In fact, there we can easily bound the outage probability as
\begin{eqnarray}
\mathcal{O}(K, \rho, r) &=& \mathrm{Pr}\left[R(K, \rho) \leq r\left|\frac{K_0}{K} \geq 1 - \epsilon\right.\right]\cdot\mathrm{Pr}\left[\frac{K_0}{K} \geq 1 - \epsilon\right]\nonumber\\
&\leq& \mathrm{Pr}\left[\frac{K_0}{K} \geq 1 - \epsilon\right];\\
\mbox{and}\;\;\mathcal{O}(K, \rho, r) &\geq& \mathrm{Pr}\left[R(K, \rho) \leq r, K_0 = K\right]\nonumber\\
&=& \mathrm{Pr}\left[R(K, \rho) \leq r| K_0 = K\right]\cdot \mathrm{Pr}[K_0 = K]\nonumber\\
&=& \mathrm{Pr}[K_0 = K],
\end{eqnarray}
by noting that, in the on-off power allocation, no positive rate can be achieved by the ``off'' parallel channels. The resulting expression of $\tilde{\mathcal{E}}_0(\eta)$ thus reduces to $\tilde{\mathcal{E}}_0(\eta) = -\log p_0 = \tau - \log(e^\tau - 1)$, as given in Proposition \ref{prop:oe-onoff}.

Case [b] ($r \geq r_c$):

Conditioning upon $K_0$ and fixing an arbitrarily small $\epsilon > 0$, we can expand the outage probability as
\begin{eqnarray}
\label{eqn:general-case-b-decomp}
\mathcal{O}(K, \rho, r) &=& \mathrm{Pr}[R(K, \rho)\leq r]\nonumber\\
&=& \mathrm{Pr}\left[R(K, \rho)\leq r \left|\frac{K_0}{K}< \epsilon\right.\right] \cdot\mathrm{Pr}\left[\frac{K_0}{K}< \epsilon\right]\nonumber\\
&&\quad\quad + \sum_{i = 1}^{1/\epsilon - 2} \mathrm{Pr}\left[R(K, \rho)\leq r \left|\frac{K_0}{K} \in [i\epsilon, (i + 1)\epsilon)\right.\right] \cdot\mathrm{Pr}\left[\frac{K_0}{K}\in [i\epsilon, (i + 1)\epsilon)\right]\nonumber\\
&&\quad\quad + \mathrm{Pr}\left[R(K, \rho)\leq r \left|\frac{K_0}{K}\geq 1 - \epsilon\right.\right] \cdot\mathrm{Pr}\left[\frac{K_0}{K} \geq 1 - \epsilon\right].
\end{eqnarray}
Therefore as $K \rightarrow \infty$, the wideband outage exponent $\mathcal{E}(\eta)$ is equal to the dominating, namely, the minimum, exponent among the $1/\epsilon$ terms in (\ref{eqn:general-case-b-decomp}).\footnote{Again, it loses no essential generality by assuming that $1/\epsilon$ is integer-valued, for sufficiently small values of $\epsilon$ as $K \rightarrow \infty$.}

To evaluate these exponential decay rates, we start with the following lemma.
\begin{lem}
\label{lem:general-case-b}
In (\ref{eqn:general-case-b-decomp}), for every fixed $x \in (0, 1)$, as $\epsilon \rightarrow 0$ with correspondingly increasing index $i$ such that $x \in [i\epsilon, (i + 1)\epsilon)$, the conditional outage probability
\begin{eqnarray*}
\mathrm{Pr}\left[R(K, \rho) \leq r \left|\frac{K_0}{K} \in [i\epsilon, (i + 1)\epsilon)\right.\right]
\end{eqnarray*}
vanishes with the following exponential decay rate:
\begin{eqnarray}
\label{eqn:ex}
\mathcal{V}_x = \sup_{\lambda \leq 0}\left\{\frac{\lambda}{\eta} - x \log\mathbf{E}\left[\exp\left(\frac{g_0\lambda}{x}A_1^{(0)}\right)\right] - (1 - x) \log\mathbf{E}\left[\exp\left(\frac{g_1 \lambda}{1 - x}A_1^{(1)}\right)\right]\right\}.
\end{eqnarray}
\end{lem}
{\it Proof of Lemma \ref{lem:general-case-b}:} We still start with bounding the conditional outage probability from above and below. In view of the expression of $R(K, \rho)$ in (\ref{eqn:rate-raw}), we have
\begin{eqnarray}
\mathrm{Pr}[R_\mathrm{u}(K, \rho) \leq r] \leq \mathrm{Pr}\left[R(K, \rho)\leq r\left|\frac{K_0}{K} \in [i\epsilon, (i + 1)\epsilon)\right.\right] \leq \mathrm{Pr}[R_\mathrm{l}(K, \rho) \leq r],
\end{eqnarray}
where
\begin{eqnarray}
R_\mathrm{u}(K, \rho) &:=& \sum_{i = 1}^{(i + 1)\epsilon K} \log\left(1 + \frac{g_0\rho}{i\epsilon K}A_i^{(0)}\right) + \sum_{j = 1}^{(1 - i\epsilon)K} \log\left(1 + \frac{g_1 \rho}{[1 - (i + 1)\epsilon]K} A_j^{(1)} \right),\\
R_\mathrm{l}(K, \rho) &:=& \sum_{i = 1}^{i\epsilon K} \log\left(1 + \frac{g_0\rho}{(i + 1)\epsilon K}A_i^{(0)}\right) + \sum_{j = 1}^{[1 - (i + 1)\epsilon]K} \log\left(1 + \frac{g_1 \rho}{(1 - i\epsilon)K} A_j^{(1)} \right)
\end{eqnarray}
bound $R(K, \rho)$ from above and below, respectively, conditioned upon $K_0/K \in [i\epsilon, (i + 1)\epsilon)$.

We then follow the large-deviation procedure in the proof of Proposition \ref{prop:general} to characterize the exponential behavior of $\mathrm{Pr}[R_\mathrm{u}(K, \rho) \leq r]$ and $\mathrm{Pr}[R_\mathrm{l}(K, \rho) \leq r]$. The corresponding asymptotic logarithmic moment generating functions are
\begin{eqnarray}
\Lambda_\mathrm{u}(\lambda) &=& (i + 1)\epsilon\log\mathbf{E}\left[\exp\left(\frac{g_0\rho\lambda}{i\epsilon} A_1^{(0)}\right)\right] + (1 - i\epsilon)\log\mathbf{E}\left[\exp\left(\frac{g_1 \rho\lambda}{1 - (i + 1)\epsilon} A_1^{(1)}\right)\right]\\
\Lambda_\mathrm{l}(\lambda) &=& i\epsilon\log\mathbf{E}\left[\exp\left(\frac{g_0\rho\lambda}{(i + 1)\epsilon} A_1^{(0)}\right)\right] + [1 - (i + 1)\epsilon]\log\mathbf{E}\left[\exp\left(\frac{g_1 \rho\lambda}{1 - i\epsilon} A_1^{(1)}\right)\right]
\end{eqnarray}
for $R_\mathrm{u}(K, \rho)$ and $R_\mathrm{l}(K, \rho)$, respectively. Consequently, the large deviation rate function of $R(K, \rho)$ can be bounded by utilizing $\Lambda_\mathrm{u}(\lambda)$ and $\Lambda_\mathrm{l}(\lambda)$, and invoking the G\"{a}tner-Ellis theorem, as in Proposition \ref{prop:general}. As we let the arbitrarily small $\epsilon > 0$ vanish asymptotically, the upper and lower bounds of the rate function of $R(K, \rho)$ asymptotically coincide, and converge to $\mathcal{V}_x$ as given by (\ref{eqn:ex}). This establishes Lemma \ref{lem:general-case-b}.

Next we characterize $\mathrm{Pr}[K_0/K \in [i\epsilon, (i + 1)\epsilon)]$. Since $K_0/K$ is the empirical mean of $K$ i.i.d. Bernoulli random variables, its large deviation rate function can be evaluated by utilizing Cram\'{e}r's theorem, as
\begin{eqnarray}
\label{eqn:ld-kb}
\inf_{x \in [i\epsilon, (i + 1)\epsilon)} \left\{x \log\frac{p_1}{p_0} - \log p_1 - H_\mathrm{b}(x)\right\}.
\end{eqnarray}
For every fixed $x \in (0, 1)$, as $\epsilon \rightarrow 0$, the infimum operator in (\ref{eqn:ld-kb}) becomes surplus and the rate function converges to $\{x\log\frac{p_1}{p_0} - \log p_1 - H_\mathrm{b}(x)\}$.

Combining Lemma \ref{lem:general-case-b} and the rate function of $\mathrm{Pr}[K_0/K \in [i\epsilon, (i + 1)\epsilon)]$, as we let the quantization step $\epsilon > 0$ vanish, we obtain the dominating exponential decay rate for the interior interval $x \in (0, 1)$ as $\inf_{x \in (0, 1)} \tilde{\mathcal{E}}(\eta, x)$, where
\begin{eqnarray}
\tilde{\mathcal{E}}(\eta, x) := \mathcal{V}_x + x \log\frac{p_1}{p_0} - \log p_1 - H_\mathrm{b}(x).
\end{eqnarray}
Using the probability density functions of $A_1^{(0)}$ and $A_1^{(1)}$, we can evaluate the expectations in $\mathcal{V}_x$ as
\begin{eqnarray}
\mathbf{E}\left[\exp\left(\frac{g_0 \lambda}{x}A_1^{(0)}\right)\right] &=& \frac{1 - e^{-(1 - g_0\lambda/x)\tau}}{p_0 (1 - g_0\lambda/x)}\\
\mathbf{E}\left[\exp\left(\frac{g_1 \lambda}{1 - x}A_1^{(1)}\right)\right] &=& \frac{e^{-[1 - g_1\lambda/(1 - x)]\tau}}{p_1 [1 - g_1\lambda/(1 - x)]}.
\end{eqnarray}
After manipulating terms, we can simplify the expression of $\tilde{\mathcal{E}}(\eta, x)$ to
\begin{eqnarray}
\tilde{\mathcal{E}}(\eta, x) := \sup_{\lambda \leq 0}\left\{\frac{\lambda}{\eta} - x \log\left[e^{(1 - g_0 \lambda/x)\tau} - 1\right] + x \log(x - g_0\lambda) + (1 - x)\log(1 - x - g_1\lambda) + (1 - \lambda)\tau
\right\}.
\end{eqnarray}

To finalize the proof, we still need to evaluate the exponential decay rates of the two extreme case terms in (\ref{eqn:general-case-b-decomp}):
\begin{eqnarray*}
\mathrm{Pr}\left[R(K, \rho)\leq r \left|\frac{K_0}{K}< \epsilon\right.\right] \cdot\mathrm{Pr}\left[\frac{K_0}{K}< \epsilon\right] \;\mbox{and}\; \mathrm{Pr}\left[R(K, \rho)\leq r \left|\frac{K_0}{K} \geq 1 - \epsilon\right.\right] \cdot\mathrm{Pr}\left[\frac{K_0}{K} \geq 1 - \epsilon\right].
\end{eqnarray*}
The exponential decay rate of the second term has already been obtained as $\tilde{\mathcal{E}}_0(\eta)$ in Case [a] of the proof. The first term can also be characterized following essentially the same procedure. We find the following results. On the one hand, the limiting point of $\tilde{\mathcal{E}}(\eta, x)$ as $x \downarrow 0$ actually coincides with the exponential decay rate of the first extreme case term above as $\epsilon \rightarrow 0$. On the other hand, the second extreme case term (which would correspond to $x = 1$) exhibits a singular behavior. That is, $\lim_{x\uparrow 1} \tilde{\mathcal{E}}(\eta, x)$ does not converge to $\tilde{\mathcal{E}}_0(\eta)$. So in order to obtain the wideband outage exponent $\mathcal{E}(\eta)$, we still need to compare $\inf_{x\in (0, 1)}\tilde{\mathcal{E}}(\eta, x)$ against $\tilde{\mathcal{E}}_0(\eta)$. This leads to the final expression of $\mathcal{E}(\eta)$ in Proposition \ref{prop:oe-general} and concludes its proof.

{\bf Q.E.D.}

\begin{figure}[ht]
\psfrag{xlabel}{{$g_0$}}
\psfrag{ylabel}{{$\tau$} (dB)}
\psfrag{zlabel}{{$\mathcal{E}(\eta)$}}
\epsfxsize=4.5in
\epsfclipon
\centerline{\epsffile{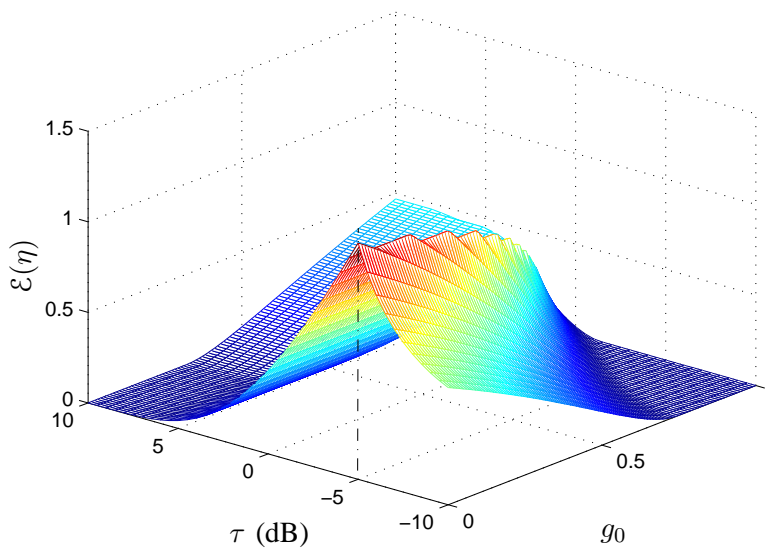}}
\caption{The mesh plot of the wideband outage exponent $\mathcal{E}(\eta)$ versus $(g_0, \tau)$, for achieving the target energy per nat $\eta = 5\mbox{dB}$.}
\label{fig:fdbkgen5db}
\end{figure}

\begin{figure}[ht]
\psfrag{xlabel}{{$g_0$}}
\psfrag{ylabel}{{$\tau$} (dB)}
\psfrag{zlabel}{{$\mathcal{E}(\eta)$}}
\epsfxsize=4.5in
\epsfclipon
\centerline{\epsffile{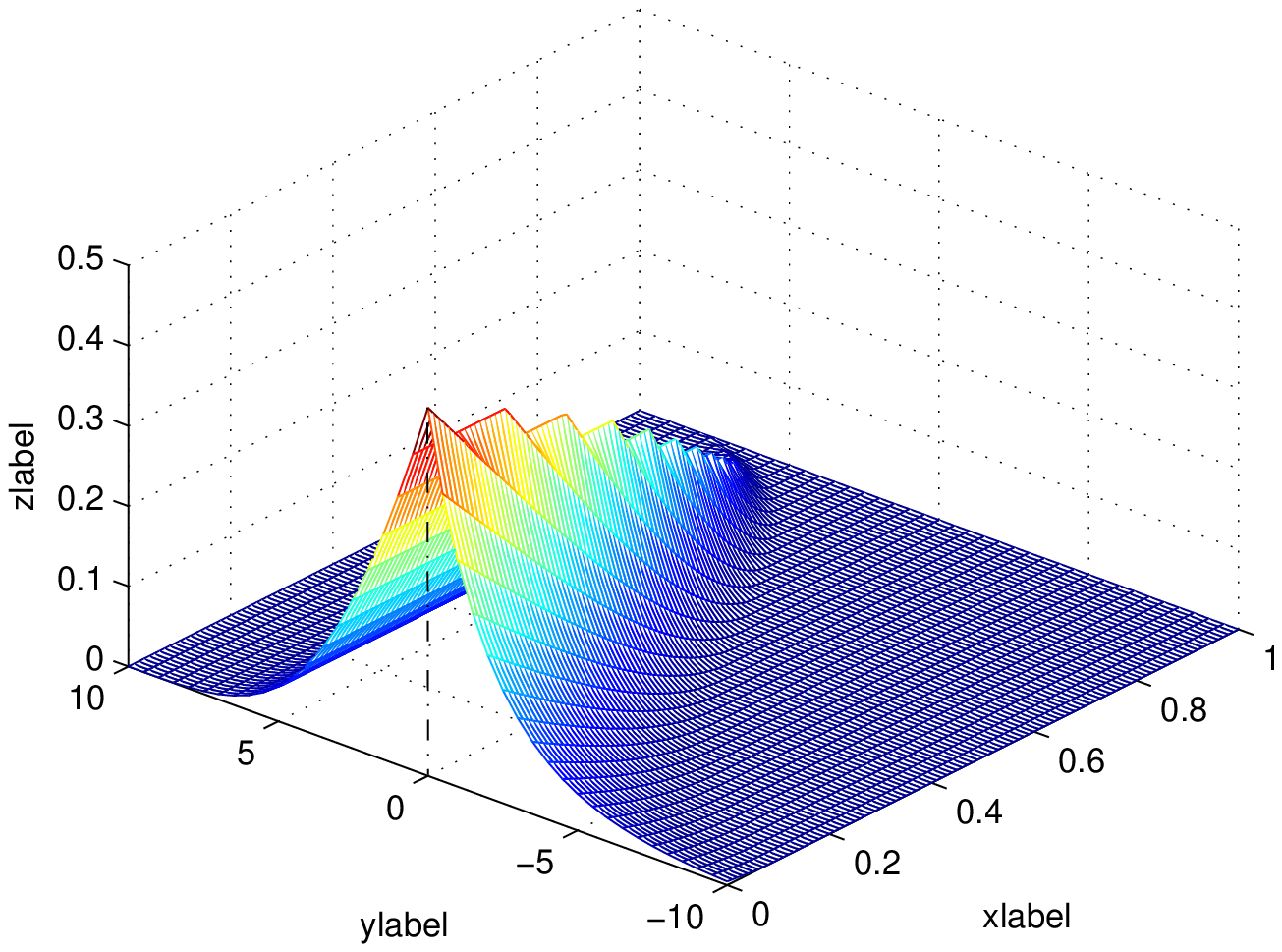}}
\caption{The mesh plot of the wideband outage exponent $\mathcal{E}(\eta)$ versus $(g_0, \tau)$, for achieving the target energy per nat $\eta = 0\mbox{dB}$.}
\label{fig:fdbkgen0db}
\end{figure}

\begin{figure}[ht]
\psfrag{xlabel}{{$g_0$}}
\psfrag{ylabel}{{$\tau$} (dB)}
\psfrag{zlabel}{{$\mathcal{E}(\eta)$}}
\epsfxsize=4.5in
\epsfclipon
\centerline{\epsffile{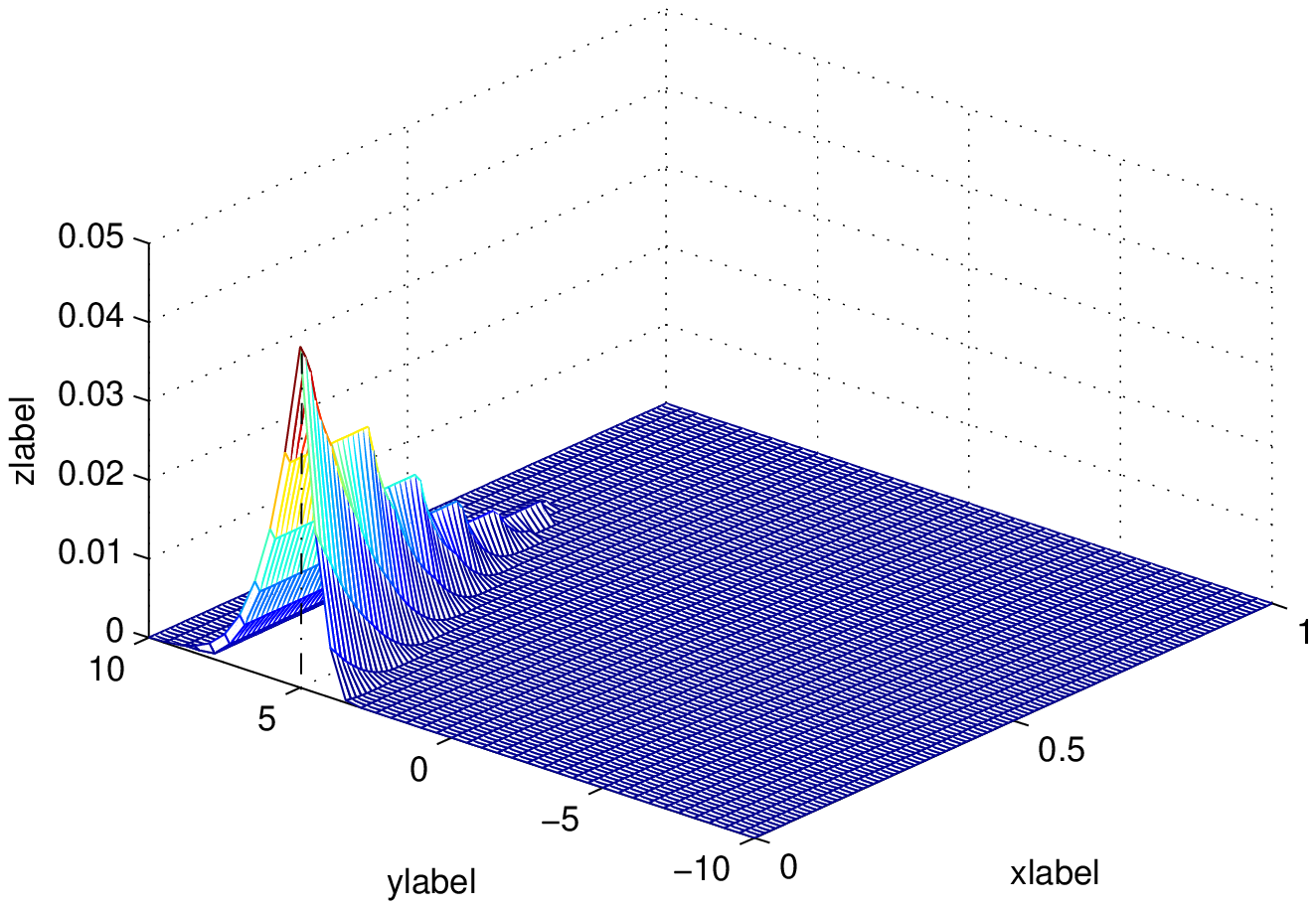}}
\caption{The mesh plot of the wideband outage exponent $\mathcal{E}(\eta)$ versus $(g_0, \tau)$, for achieving the target energy per nat $\eta = -5\mbox{dB}$.}
\label{fig:fdbkgenm5db}
\end{figure}

For a given target energy per nat $\eta > 0$ and a given transmission protocol, we can then utilized Proposition \ref{prop:oe-general} to compute the resulting wideband outage exponent $\mathcal{E}(\eta)$. Furthermore, by optimizing over the protocol parameters $\tau$ and $g_0$, we can compute the maximum achievable $\mathcal{E}(\eta)$. Figures \ref{fig:fdbkgen5db}-\ref{fig:fdbkgenm5db} display the mesh plots of $\mathcal{E}(\eta)$ versus $(g_0, \tau)$, for $\eta = 5\mbox{dB}, 0\mbox{dB}$, and $-5\mbox{dB}$, respectively. Interestingly, all the three plots numerically show that the maximum $\mathcal{E}(\eta)$ is achieved by the optimal on-off power allocation scheme with $g_0 = 0$ and $\tau = 1/\eta$. Although we still lack an analytical proof for this claim, it is tempting to conjecture that the on-off power allocation scheme with $\tau = 1/\eta$ maximizes the wideband outage exponent among all the transmission protocols as described in Section \ref{subsec:protocol}.

From Figures \ref{fig:fdbkgen5db}-\ref{fig:fdbkgenm5db} we further notice that, $\mathcal{E}(\eta)$ typically appears like a ``ridge'' decreasing in a piecewise constant way along $g_0$. Thus we shall expect that slightly changing the power allocation between the two-level channel states does not drastically alter the wideband outage exponent. In contrast, the ``ridge'' rapidly rolls off along $\tau$, implying that the wideband outage exponent is rather sensitive with respect to even small changes in the CSI quantization level.

\subsection{Further Discussions of Partial Channel State Feedback}

It is essential to note that, in this section we have only investigated one specific transmission protocol with one-bit channel state feedback for each parallel channel. There can be various transmission protocols other than what we considered, and it is unclear which of them yields the optimal performance. In this subsection we briefly outline two such alternative transmission protocols.

First, instead of marking each parallel channel based upon whether its fading level exceeds $\tau$, we can sort the $K$ parallel channels based upon their fading levels, and mark the $K^\prime$ parallel channels with the highest fading levels. Intuitively, such a feedback strategy appears to yield more robust performance than the transmission protocol we considered, because it ensures the existence of a constant number ({\it i.e.}, $K^\prime$) of ``good'' parallel channels. But on the other hand, these marked ``good'' parallel channels may in fact have rather low fading levels. So it is unclear whether this alternative strategy performs better or poorer than the transmission protocol we considered. The difficulty of analyzing this strategy is that the $K^\prime$ ``good'' parallel channels are no longer independent, because their joint distribution is obtained by partially marginalizing the ordered statistics of the $K$ i.i.d. fading levels.

Another alternative transmission protocol can be obtained by allowing variable bit-allocation in feeding back the $K$ parallel channel states. Under the total feedback rate constraint of $K$ bits, we may allow a fraction of parallel channels to quantize their fading levels using more than one single bit, while force another fraction of parallel channels not to feed back any information about their fading levels. In this way, the transmitter is able to identify some particularly strong parallel channels, at the cost of totally losing information about some other potentially strong parallel channels. Hence there exists some tradeoff affecting the resulting wideband outage exponent remaining to be explored.

\section{Conclusion}
\label{sec:conclusion}

Motivated by wideband communication in certain slow-fading propagation environments such as indoor systems, we investigate a mathematical model of wideband quasi-static fading channels, focusing on its asymptotic behavior of channel outage probability in the wideband regime. This work is complementary to the vast body of prior research on ergodic wideband channels, which are more suitable for modeling long-term system throughputs in time-varying propagtion environments.

The main finding in this paper essentially reiterates a simple theme that, a communication system is unlikely to simultaneously maintain both high rate and high reliability, or, both high energy efficiency and high spectral efficiency. As the target energy efficiency approaches the minimum achievable energy per nat in the wideband limit, the outage probability increases, as quantitatively captured by the wideband outage exponent. Such tradeoffs can be leveraged by several factors, like channel fading models, deployment of multiple antennas, antenna spatial correlation, and partial transmit knowledge of the channel state information. In this paper we illustrate the effect of these factors through a series of representative case studies, and the accumulated insights may serve as useful guidelines for performing first-cut analysis of wideband transmission systems.

\bibliographystyle{ieee}
\bibliography{./wbout}

\end{document}